\documentclass[prd,amsfonts,amssymb,amsmath,graphicx,epsfig]{revtex4}
\usepackage{epsfig}
\date{\today}
\begin{document}

\title{Gravity theories with local energy-momentum exchange: \\
a closer look at 
%rescuing 
Rastall-like gravity}
%\title{Why is gravity oblivious to the ultraviolet divergence of quantum theory?}
\author{Daniel A.\ Turolla Vanzella}\email{vanzella@ifsc.usp.br}
\affiliation{Instituto de F\'\i sica de S\~ao Carlos,
Universidade de S\~ao Paulo, Caixa Postal 369, CEP 13560-970, 
S\~ao Carlos, SP,  Brazil}

\begin{abstract}
Einstein's famous equivalence principle
is certainly one of the most striking
 features of the gravitational interaction. In a strict reading, 
 it states that  the effects of gravity can be made to disappear
{\it locally} by a convenient choice of reference frame.
As a consequence, no covariantly-defined gravitational force
should exist and energy-momentum of all matter and interaction fields combined,
with gravity {\it excluded}, should be locally conserved. Although elegant,
this separate conservation law represents a strong constraint on the dynamics
of a gravitating system and it is only logical to question its naturality and
observational basis. This is the purpose of the present work. For concreteness sake,
we analyze, in the context of metric theories of gravity, 
the simplest phenomenological model which allows for local energy-momentum exchange between the
spacetime and matter/interaction fields while preserving 
the seemingly more
natural principle of general covariance.
%It turns out that
%our concrete model can be seen as a generalization
This concrete model turns out to be a generalization
of the socalled Rastall's  theory, with one important
advantage: criticisms 
%commonly 
made to the latter, which are often 
used to dismiss it as a viable or interesting model, 
do not apply to 
the former in a universe containing ``dark'' ingredients,
such as ours ---
a connection  which seems to have been overlooked 
thus far. 
%In particular,
%by imposing compatibility with 
%Newtonian gravity in the proper 
%regime,
%we explicitly show that 
%
%widespread beliefs --- such as (i)~that 
%Rastall's theory is completely equivalent to General Relativity and
%(ii)~that its interpretation as a nonconservative gravity theory cannot 
%be made consistent with flat-space physics --- are based on
%arguments which are evaded in a Universe containing ``dark'' ingredients, such as ours --- a connection  which seems to have been overlooked 
%thus far. 
%by imposing compatibility with Newtonian gravity in the proper 
%regime, 
%this issue of ``nonconservative gravity'' is inevitably tied to the 
%existence
%of ``dark'' ingredients --- a connection  which seems to have been overlooked 
%thus far. 
We conclude by exploring 
%We explore 
the
consequences of our Rastall-like model 
to standard  (astrophysical and 
cosmological) gravitational scenarios.

\end{abstract}

%\pacs{04.62.+v}

\maketitle
\section{Introduction}
\label{sec:intro}

Local energy-momentum conservation has become one of the cornerstones upon which we lay the foundations of our physical theories.
In its most general form, it is expressed by the vanishing of the 4-divergence of the stress-energy-momentum tensor $T^{ab}$
of an isolated system: $\nabla_a T^{ab}=0$ --- where $\nabla_a$ is the covariant derivative operator compatible with the spacetime metric
$g_{ab}$. In words, this equation captures the idea  of energy-momentum (i.e., 4-momentum) of a system 
not being (covariantly) 
created nor destroyed anywhere {\it locally} --- which we may take as the definition of an isolated system. 
In particular,
it is usually assumed that the system comprised by {\it all} matter and interaction fields combined,  with gravity  {\it excluded}, constitutes an isolated system
{\it per se}. This assumption finds its roots in a strict reading of the
famous Einstein's  
equivalence principle (EEP), according to which the effects 
of gravity on the dynamics of a system can be made to disappear {\it locally}. In fact, many 
metric theories of gravity enforce this through their
field equations,
\begin{eqnarray}
{\cal G}^{ab} = \kappa T^{ab},
\label{eq:metrictheories}
\end{eqnarray}
with $\kappa$ being a ``coupling constant'' and 
where ${\cal G}^{ab}$ is a tensor obtained from the functional derivative (w.r.t.~the metric $g_{ab}$) of some  action built out of a
geometric Lagrangian scalar density ${\cal L}_g$ --- General Relativity (GR) 
being the special case where ${\cal L}_g \propto R$ (the Ricci scalar curvature)
and ${\cal G}^{ab} = G^{ab}$ (the Einstein's tensor). The geometric identity $\nabla_a{\cal G}^{ab} \equiv 0$ --- true for any ${\cal G}^{ab}$ obtained
in the way described above --- makes local energy-momentum conservation a direct consequence of gravity. In this framework, whatever
pervades the spacetime, it influences and gets influenced by the latter
%, through Eq.~(\ref{eq:metrictheories}), 
without ever exchanging 4-momentum locally with it.

The idea of having local energy-momentum conservation enforced by gravity itself is undeniably elegant and simple and it played an
important role in  arriving at the theory of GR. However, it is only logical to question its naturality and, most importantly, how 
tightly constrained it is by the observational data. This was the
the sole  purpose of the present work. 
Similar investigations have been conducted in the 
literature (see Ref.~\cite{Rev} and references therein).
For instance, metric theories of gravity violating 
$\nabla_aT^{ab}= 0$ can be obtained, through a 
variational principle, in the scope of 
$f(R,T)$ theories~\cite{HLNO} --- in which
a Lagrangian density given by $f(R,T)$ is postulated ($T$ being 
the trace of $T^{ab}$). In the absence of a fundamental ``microscopic''
description of 
energy-momentum exchange between the spacetime and 
matter/interaction fields,
here we take a phenomenological approach and focus directly on
the effective field equations as starting point, 
not worrying  whether or not there
is a correspondent Lagrangian formulation. 
In particular, 
the concrete
analysis conducted here is based on a
generalization of Einstein's equations
%which is implied by the nonconservative equation
 --- implied by 
Eqs.~(\ref{model1}) and (\ref{model1j}) below ---
which allows for 
energy-momentum exchange between the spacetime and 
matter/interaction fields, while 
preserving the seemingly more natural principle of general
covariance.

Before we proceed, we must make full disclosure that,
after completing the writing of the present manuscript, 
it came to our attention 
that a {\it particular} realization of our concrete model 
%given 
%by Eq.~(\ref{model1})
%and (\ref{model1j})  
%have 
has been  analyzed 
in the literature under the name of 
Rastall's gravity~\cite{Rastall,LH,BDFPR,OVFC,OVF,Visser}. 
In Rastall's model, the parameter $\sigma$
appearing in Eq.~(\ref{model1}) 
%(or, equivalently, the socalled
%Rastall's parameter $\lambda = c^4 \sigma/(8\pi G_N)$) 
is taken to be universal, in the sense that its value is
independent of the matter/field content in the spacetime.
Although this may seem to be a minor detail, it turns out that
it is enough to make Rastall's proposal vulnerable
to criticisms which have been used to dismiss it as a viable or interesting model~\cite{LH,Visser}.
%incompatible with flat-space
%physics~\cite{LH} 
%and with Newtonian gravity (unless $\sigma$ is completely 
%negligible). 
In sharp contrast, by not making any universality
assumption, our Rastall-like model can avoid this 
shortcoming provided we make one interesting concession:
existence of ``dark'' ingredients (i.e.,
constituents which we  perceive only through their
gravity effects) --- a topic 
which was mainly dormant when Rastall proposed his model~\cite{Rastall}.
%%
%
%
%
%This might 
%seem to render the present
%analysis
%redundant, but we believe 
%this is not  the case. To the best of our knowledge
%and understanding, Rastall's original analysis and  subsequent
%work based on it lack 
%one key point when it comes to confronting 
%their implications with observations: consistency with the
%Newtonian-gravity regime.  
%In particular,
%consistency with Newtonian gravity in the proper regime
%inevitably ties our model (in ``nonconservative gravity'' as a whole) 
%to
%the existence of ``dark'' ingredients in our Universe --- i.e.,
%constituents which we  perceive only through their
%gravity effects ---, a topic 
%which was mainly dormant when Rastall proposed his model~\cite{Rastall}. 
Although
Rastall's theory has been later considered in the cosmological  context 
of 
dark energy
and dark matter~\cite{BDFPR}, this was mainly done 
as if they were
independent subjects, missing or overlooking their 
codependency.

Having the disclosure above been made,
%--- clearly waiving any claim 
%of primacy
%w.r.t.~the model given in Eq.~(\ref{model1}) and its motivation ---, 
we have opted to
keep our presentation
as self-contained as possible,
making 
it easier for the reader to follow the analysis.
In Sec.~\ref{sec:dynst}, we revisit the 
case for investigating the possibility of energy-momentum 
being exchanged between
the spacetime and matter/interaction fields. In
Sec.~\ref{sec:model},
we focus on 
a simple phenomenological model in which energy-momentum exchange
can occur, without violating general covariance
--- which turns out to be a generalization of 
the  one proposed by Rastall fifty years ago~\cite{Rastall}.
In Sec.~\ref{sec:modeq}, we  discuss in detail
how criticisms used to dismiss the original Rastall's model 
do not pose any serious challenge to our proposal.
In the process, we also obtain the modified Einstein's equations 
which are 
consistent with the model presented in the previous section 
and, in 
addition, which recover
Newtonian gravity in the proper regime. 
In particular, it is shown that a matter/field constituent 
which
exchanges energy-momentum with the spacetime in a 
nonnegligible manner 
(called ``nonconservative,'' for short)
cannot exert negligible pressure $P$
(compared to its energy density $\rho$) when considering
inhomogeneous energy distributions, even in the
Newtonian regime. This result inevitably ties our 
investigation to the
existence of ``dark'' ingredients, 
a twist which was not intended by our primary motivation.
In Sec.~\ref{sec:NRw}, we discuss in more detail 
the 
%nonrelativistic behavior
Newtonian regime
of nonconservative ingredients, while in Sec.~\ref{sec:cc} we analyze the
specific case of a cosmological-constant-like ($P=-\rho$)
nonconservative component. Finally, in Secs.~\ref{sec:modSch} 
and \ref{sec:modcosm}, we apply the modified
Einstein's equations to standard (astrophysical and
cosmological) scenarios: (i) static, spherically symmetric
and (ii) spatially homogeneous and isotropic ones. 
%Many of the results of these latter sections,
%summarized in Figs.~\ref{fig:w0m1by3diag}-\ref{fig:diagsw0},
%depend crucially on
%the Newtonian-limit analysis of Sec.~\ref{sec:modeq},
%constituting original contributions to the topic of Rastall's 
%theory.
Sec.~\ref{sec:final} is dedicated to our concluding remarks.

\section{Minkowski vs.~dynamical spacetimes}
\label{sec:dynst}

In the context of special relativity, local (leading to global) 
energy-momentum conservation for the system ${\cal U}$ comprising {\it all} matter and interaction fields
(the ``${\cal U}$niverse'') is a quite natural imposition since the background Minkowski spacetime
has no dynamical degrees of freedom; ${\cal U}$ is naturally an isolated system. In contrast to that, metric theories of gravity
turn the spacetime 
degrees of freedom on, letting ${\cal U}$ interact with them. Now, the assumption that such an interaction should not
transfer 4-momentum to/from the system ${\cal U}$ is far from obvious and represents a strong (although elegant) constraint on the dynamics 
of ${\cal U}$. In fact, one can think of several ways in which 
${\cal U}$ and the spacetime might exchange 4-momentum, even without ever violating the seemingly more natural principle of
general covariance --- according to which the spacetime metric $g_{ab}$
is the only quantity external to the system ${\cal U}$ which can influence its dynamics.
%(which seems to be a more general principle). 
The simplest such example, proposed originally by Rastall~\cite{Rastall}, is 
$\nabla_b T^{ab} = \lambda \nabla^aR$, with $\lambda$  some constant with dimension of $c^4/G_N$ ($c$ being the speed of light in 
vacuum and $G_N$ the Newton's constant).

In the absence of a description of some fundamental 
%description of 
mechanism through which 4-momentum could be exchanged between
the spacetime and matter/interaction fields, the merit of such speculation
should be evaluated from compliance with well-established general principles and by
confronting concrete (phenomenological) proposals with observations and experimental constraints.
Here, we analyze one such phenomenological 
model which can be seen as a generalization
of Rastall's 
proposal, simultaneously preserving general covariance and the EEP in the regime where it has been
convincingly tested --- namely, the
test-particle regime in the vacuum ---, and we also constrain the final theory to have the proper weak-gravity Newtonian limit --- which ties this subject of ``nonconservative'' gravity to the existence of
``dark'' ingredients.

\section{The minimal (Rastall-like) model}
\label{sec:model}

As mentioned in the previous section, we shall investigate the consequences of the simplest model 
in which  4-momentum can be exchanged between the spacetime and matter/interaction fields (with stress-energy-momentum tensor $T^{ab}$):
\begin{eqnarray}
\nabla_b T^{ab} = \frac{c^4 \sigma}{8 \pi G_N} \nabla^aR,
\label{model1}
\end{eqnarray}
where $\sigma$ is a dimensionless constant 
(with the pre-factor $c^4/(8\pi 
G_N)$ chosen only for  convenience). One might take different approaches
toward this constant. For instance,
one might want to consider it as fundamental and universal --- in  the sense that 
$\sigma$ would be independent of (the constituents of) 
${\cal U}$. This would take us directly to Rastall's original 
proposal~\cite{Rastall}.
However, this view is subject to 
two major criticisms. First, it has been argued~\cite{LH,Visser} that 
Rastall's theory is completely equivalent to standard (i.e.,
``conservative'') 
GR. The reason is 
that $T^{ab}$ satisfying Eq.~(\ref{model1}) would {\it not} 
be recognized as the  ``physical'' stress-energy-momentum tensor;
instead, a {\it conserved} tensor ${\cal T}^{ab}$ (constructed from 
$T^{ab}$ and its trace) can be defined, in terms of which the gravity field
equations would reduce to Einstein's equations with ${\cal T}^{ab}$ playing the role of
the physical stress-energy-momentum tensor. The second criticism is that, even if one concedes that
$T^{ab}$ appearing in Eq.~(\ref{model1}) might be the physical stress-energy-momentum tensor,
consistency with
flat-space physics would force $\sigma$ to be completely negligible, rendering Rastall's theory
irrelevant
for all practical purposes.
%We shall discuss these criticisms in more detail in Sec.~\ref{sec:modeq}.

%inconsistent with our observed
%Universe unless $\sigma$ is completely negligible, as 
%has been pointed out in the literature~\cite{LH} and we 
%shall see later in detail.
Here, we take a different approach, which is to consider that each matter/interaction-field constituent 
${\cal U}_j$ of ${\cal U}$ has its own effective constant $\sigma_j$:
\begin{eqnarray}
\nabla_b T_{(j)}^{ab} = \frac{c^4 \sigma_j}{8 \pi G_N} \nabla^aR + f^a_{(j)},
\label{model1j}
\end{eqnarray}
where $T_{(j)}^{ab}$ is the stress-energy-momentum tensor of the subsystem ${\cal U}_j$ and $f^a_{(j)}$ is the ``external'' 4-force density acting on
${\cal U}_j$ due to ``direct'' 
interaction (i.e., not mediated by the spacetime geometry) 
with $\{{\cal U}_k\}_{k\neq j}$ --- so that $\sum_j f^a_{(j)} = 0$. In this latter scenario, summing up the contribution of Eq.~(\ref{model1j}) for each constituent
leads to Eq.~(\ref{model1})
with $\sigma = \sum_{j}\sigma_j$. It should be stressed, though, that the set of
Eqs.~(\ref{model1j}) for ${\cal U} = \{{\cal U}_j\}$ (together with the
gravity field equations)
is to be seen as more fundamental than 
Eq.~(\ref{model1}), so that
%Therefore, 
any reasoning
about what the ``physical'' stress-energy-momentum tensor of ${\cal U}_j$ looks like, 
in the lines presented in Refs.~\cite{LH,Visser}, should be based on the former (see Sec.~\ref{sec:modeq}).
%Interestingly enough, 
%As we shall see below,
%this leads to
%a Rastall-like
%model which evades all criticisms which plague Rastall's original 
%proposal. 

%\footnote{
%In this latter approach, i
Before moving further, it is 
%also 
interesting to speculate that in case this 4-momentum transfer is quantum mechanical in nature 
--- just because this is the least-understood regime of gravity ---, 
a natural scale for $\lambda_j := c^4\sigma_j/(8\pi G_N)\sim Energy/length$ could be $m_jc^2/[\hbar/(m_jc) ] = m_j^2c^3/\hbar$, where $m_j$ 
is the mass of the matter/interaction 
field being considered (and $\hbar$ is the reduced Planck's constant). 
This would make $(m_j/m_P)^2$ the natural scale for $\sigma_j$ ($m_P :=\left(\hbar c/G_N\right)^{1/2}$ being the Planck mass),
rendering this 4-momentum exchange currently negligible for all {\it known} fields but possibly relevant for a hypothetical ultra-massive field --- which might presumably 
be in its vacuum state today.

\section{Modified Einstein's equations, nonequivalence with General Relativity and the Newtonian-gravity regime}
\label{sec:modeq}

Since our approach here is more of giving a ``proof of concept'' than exhausting all possible gravity theories with
4-momentum exchange, we focus on the case where standard GR would be recovered for $\sigma \to 0$
--- instead of any other conservative gravity theory.
The reader might think 
that this is achieved by simply adding a term $\sigma g^{ab} R$ to the left-hand side of 
Einstein's equations (Eq.~(\ref{eq:metrictheories}) with ${\cal G}^{ab} = G^{ab} := R^{ab}-g^{ab}R/2$ and $\kappa = 8\pi G_N/c^4$).
However, this is not necessarily correct:
adding a term $\sigma g^{ab} R$ to the left-hand side of Einstein's equations
is {\it not} 
the only way to implement Eq.~(\ref{model1}) and, more importantly, it 
may not 
lead to the proper weak-gravity, Newtonian limit for a given
$\sigma \neq 0$.
For this reason, we  start from the general form
\begin{eqnarray}
\alpha R^{ab} +\beta g^{ab} R = \frac{8\pi G_N}{c^4} T^{ab},
\label{modEqGen}
\end{eqnarray}
where
$T^{ab}= \sum_j T^{ab}_{(j)}$ and
$\alpha$ and $\beta$ are constants whose values depend on $\sigma$, with $\alpha \to 1$ and $\beta \to -1/2$ as $\sigma \to 0$.
Then, 
by taking the covariant derivative of Eq.~(\ref{modEqGen}), using Bianchi identity, and imposing Eq.~(\ref{model1}), we obtain $\beta = \sigma -\alpha/2$:
\begin{eqnarray}
\alpha G^{ab} +\sigma g^{ab} R = \frac{8\pi G_N}{c^4} T^{ab}.
\label{modEqGen2}
\end{eqnarray}
Note that unless $\alpha(\sigma) \equiv 1$,
%--- which is the condition usually assumed in the
%literature on Rastall's theory ---, 
Eq.~(\ref{modEqGen2}) is {\it not} equivalent to simply adding a term $\sigma g^{ab} R$ to the left-hand side of Einstein's equations, as it will become clear below.
The value of $\alpha$ for a given $\sigma$ must be fixed by imposing the correct Newtonian limit (since  $G_N$ is the observed
Newton's constant).
But before that, note that
we can rewrite Eq.~(\ref{modEqGen2}) in the equivalent forms (provided the reasonable assumption $\alpha \neq 0$)
\begin{eqnarray}
G^{ab} = \frac{8\pi G_N}{\alpha c^4}\left[
T^{ab}
+\frac{\bar{\sigma} g^{ab} T}{(1-4\bar{\sigma})}
\right]
\label{modEqGab}
\end{eqnarray}
and
\begin{eqnarray}
R^{ab} = 
\frac{8\pi G_N}{\alpha c^4}\left[
T^{ab}
+\frac{(\bar{\sigma}-1/2) }{(1-4\bar{\sigma})}g^{ab} T
\right],
\label{modEqRab}
\end{eqnarray}
where $\bar\sigma := \sigma/\alpha$ and we also have to impose the  condition $\bar\sigma \neq 1/4$ 
(i.e., $\alpha \neq 4\sigma$) --- otherwise, Eq.~(\ref{modEqGen2})
would lead to $T\equiv 0$, which seems too restrictive as a general physical condition. We promptly 
see that these modified equations
possess the same vacuum solutions 
as those of standard GR: $R^{ab} = 0$. 
%In particular, note that there is no 4-momentum exchange 
%in pure vacuum, which can be seen as an inner consistency of the model.~\footnote{This by no means implies 
%that these vacuum solutions will have the same stability
%properties as in GR. In fact, depending on the value of $\sigma$, it seems
%plausible that Eq.~(\ref{model1}) may lead to unstable vacuum solutions 
%which might favor spontaneous stress-energy-momentum ``creation''~\cite{OVF}.
%But this stability analysis, which is already intricate in GR, is beyond the scope of the present work.}

\subsection{GR, or not GR -- that is the question}
\label{subsec:NotEquiv}

%A cautionary note should be made at this point.
%w.r.t.\ Eq.~(\ref{modEqGab}). 
This is a good point to address, in the context of the present model, one  criticism which is
used to dismiss Rastall's theory: the criticism of equivalence with GR.
The reader might be tempted
to conclude, in analogy with Refs.~\cite{LH,Visser} 
regarding Rastall's original proposal, 
that Eq.~(\ref{modEqGab}) implies that 
our model is equivalent to standard GR. This reasoning  assumes that one is allowed to freely
identify whatever appears in
the r.h.s.\ of
Eq.~(\ref{modEqGab}) as (proportional to) the  (total) ``physical'' stress-energy-momentum tensor.
%This should not be confused, though, with the claim that
%these Rastall-like models and GR are physically equivalent~\cite{LH,Visser}. 
This, however, is not necessarily true. First,
note that {\it any} metric theory of gravity
can be put in the form $G^{ab} = 8\pi G_N {\cal T}^{ab}/c^4$, with 
${\cal T}^{ab}$ some tensor satisfying $\nabla_a{\cal T}^{ab}=0$ --- 
simply solve whatever the gravity field equations are, calculate
$G^{ab}$ from the solution, and then extract ${\cal T}^{ab}$ out of 
$ {\cal T}^{ab} = c^4 G^{ab}/(8\pi G_N)$. This trivial fact only means that one can always interpret 
the spacetime evolution through the lens of GR (i.e., taking GR for granted) --- possibly at the expense 
of having to
postulate ``exotic'' constituents with  {\it ad hoc},
effective
energy-momentum distributions, dynamics, and interactions.
Hence, arguing that whatever ${\cal T}^{ab}$ satisfying 
$G^{ab} = 8\pi G_N {\cal T}^{ab}/c^4$
should be seen as the physical stress-energy-momentum tensor would reduce GR to 
a mere tautology.

The point is that
although Eq.~(\ref{modEqGab}) does imply Eq.~(\ref{model1}), it is {\it not}
equivalent to (nor imply) the set of equations given by Eq.~(\ref{model1j}), which, in the present model, is more
fundamental than Eq.~(\ref{model1}).
Substituting $R$, obtained from any of the 
Eqs.~(\ref{modEqGen2})-(\ref{modEqRab}),
directly into Eq.~(\ref{model1j})
leads to
\begin{eqnarray}
%\nabla_b\left[T^{ab}_{(j)}+\frac{\bar{\sigma}_j}{(1-4\bar{\sigma})}g^{ab} %T_{(j)}\right] = -\frac{\bar{\sigma}_j}{(1-4\bar{\sigma})}
%\nabla^aT_{(\backslash{\!\!\!j})}+f^a_{(j)},
\nabla_b\left[T^{ab}_{(j)}+\frac{\bar{\sigma}_j}{(1-4\bar{\sigma})}g^{ab} T\right] = f^a_{(j)},
\label{nablacalTabj}
\end{eqnarray}
where $\bar{\sigma}_j := \sigma_j/\alpha$.
%Therefore, 
%the criticism of equivalence with GR  will apply to our model {\it only
%if}, in terms of ${\cal U}_j$'s ``physical'' stress-energy-momentum tensor
%${\cal T}^{ab}_{(j)}$ (yet to be discussed), the
%expression  above can be put
%in the form $\nabla_b {\cal T}_{(j)}^{ab} = {\cal F}^a_{(j)}$,
%with ${\cal T}^{ab} = \sum_j{\cal T}_{(j)}^{ab}$ matching the
%expression in brackets in Eq.~(\ref{modEqGab}). Is this the case?
%
%It turns out that this question, regarding ``what'' ${\cal T}_{(j)}^{ab}$ is,  {\it cannot} be tackled without
%further considerations. Like any other conservation-like equation, 
It is important to note that 
Eq.~(\ref{nablacalTabj}) does {\it not}, 
in general,
determine
the evolution of ${\cal U}_j$; it is rather a constraint,  either
enforced by
${\cal U}_j$'s own
equations of motion or which must be supplemented by additional
relations (e.g., constitutive relations, equations of state, etc.)~in 
order to determine ${\cal U}_j$'s evolution. Either way, 
additional 
equations, relating the components of 
${\cal U}_j$'s stress-energy-momentum tensor
%${\cal T}^{ab}_{(j)}$
among themselves and/or 
to ${\cal U}_j$'s kinematic
variables, {\it must} be provided. These equations, supposed to be generic ---
in the sense that they are valid for generic ${\cal U}_j$'s configurations ---
and to  involve only ${\cal U}_j$'s variables, are essential to giving 
physical/observable meaning
to ${\cal U}_j$'s 
stresses, momenta, and energies. Note that 
this already  prevents 
%${\cal T}^{ab}_{(j)}$ from being given by 
the
expression in brackets in Eq.~(\ref{nablacalTabj}) to play the role
of the physical stress-energy-momentum tensor of ${\cal U}_j$, 
since it is
$T =\sum_j T_{(j)}$, not $T_{(j)}$ alone, which appears there.
%In summary,  
Any reasoning about whether or not 
$T^{ab}_{(j)}$ is
the physical stress-energy-momentum tensor of ${\cal U}_j$, without 
addressing these additional relations,
is meaningless. 

Obviously, the whole point of the present work is 
to investigate the consequences of having Eq.~(\ref{model1j})
satisfied by the {\it physical} stress-energy-momentum tensor 
%(i.e., that ${\cal T}^{ab}_{(j)} \equiv T^{ab}_{(j)}$ --- 
(something which
Ref.~\cite{LH} recognizes to be {\it possible} even w.r.t.~Rastall's 
original proposal).
This means 
that the additional equations needed to determine ${\cal U}_j$'s evolution 
are supposed to take generic, definite forms (which may vary for different $j$) 
when expressed in terms 
of $T^{ab}_{(j)}$'s components (in a given basis). This  not
only is perfectly plausible, 
as it
is particularly appealing in cases where $T^{ab}_{(j)}$ is obtained from
general physical arguments in flat spacetime, since the geometric
nonconservative
term in Eq.~(\ref{model1j}) vanishes in that case. 
Nonetheless, 
the experimentalist/phenomenologist may well
object that equations of state
and constitutive relations are often obtained through tabulated 
experimental data
instead of theoretical modeling. In that case,
Eq.~(\ref{nablacalTabj}) may be put in a more appealing form,
\begin{eqnarray}
\nabla_b\left[T^{ab}_{(j)}+\frac{\bar{\sigma}_j}{(1-4\bar{\sigma})}g^{ab} T_{(j)}\right] = -\frac{\bar{\sigma}_j}{(1-4\bar{\sigma})}
\nabla^aT_{(\backslash{\!\!\!j})}+f^a_{(j)}
\label{nablacalTabj2}
\end{eqnarray}
(where quantities with subscript $\backslash{\!\!\!j}$ stand for their
sum over $k\neq j$),
and 
%the experimentalist/phenomenologist 
one may argue
that even if $T^{ab}_{(j)}$ is the ``true'' stress-energy-momentum tensor,
mechanical
{\it nongravitational} experiments conducted in the flat-spacetime 
{\it limit} (i.e., arbitrarily close, but not equal, to flat) might
hide $T^{ab}_{(j)}$ behind the combination
\begin{eqnarray}
{\cal T}^{ab}_{(j)} := 
T^{ab}_{(j)}+\frac{\bar{\sigma}_j}{(1-4\bar{\sigma})}g^{ab} T_{(j)},
\label{calTj}
\end{eqnarray}
which  also depends only on ${\cal U}_j$ (assuming $T_{(j)}^{ab}$ does) and, in addition, satisfies
$\nabla_b{\cal T}_{(j)}^{ab} = 0$ wherever only ${\cal U}_j$ is present
--- a situation the experimentalist, neglecting the spacetime degrees of
freedom, might  take for granted
${\cal U}_j$ to be isolated.
Note, however, that even in this case our {\it multicomponent} Rastall-like model would
{\it not} be equivalent to GR, since, in terms of
${\cal T}^{ab}_{(j)}$, Eq.~(\ref{modEqGab}) would take the form
\begin{eqnarray}
G^{ab} = \frac{8\pi G_N}{\alpha c^4}\left[
{\cal T}^{ab}
+
g^{ab}\sum_j\frac{\bar{\sigma}_{\backslash{\!\!\!j}}
{\cal T}_{(j)}}{(1-4\bar{\sigma}_{\backslash{\!\!\!j}})}
\right],
\label{modEqGabcalT}
\end{eqnarray}
with ${\cal T}^{ab}:=\sum_j{\cal T}^{ab}_{(j)}$.

In summary, we have shown not only that it is possible, but also that it
is quite 
reasonable and natural to consider that the Rastall-like model 
presented here  does {\it not}
reduce to mere GR; Eq.~(\ref{modEqGab}) 
%--- or Eq.~(\ref{modEqGabcalT}) ---
 with a given modeling for the constituents
$\{{\cal U}_j\}$ (either theoretical or phenomenological) 
leads to a {\it different} overall evolution than would GR with the
same constituents
%; 
%our 
%multicomponent 
%Rastall-like model is {\it not} equivalent
%to GR 
--- as our concrete calculations will make explicit later on.

\subsection{Newtonian-gravity regime and the modified Einstein's equations}
\label{subsec:newton}

Now, let us  turn our attention to the other criticism which is used to dismiss
Rastall's theory, namely, that it would be inconsistent with flat-spacetime physics (more precisely, with the
flat-spacetime {\it limit}) if 
$T^{ab}$ appearing in Eq.~(\ref{model1}) is taken to be the physical stress-energy-momentum tensor~\cite{LH}.

%Back to our model, 
In standard GR, 
the Newtonian limit is taken to describe nonrelativistic test particles freely falling in
``weak'' gravitational fields 
%--- in the sense that the field equations are
%linear in the metric perturbations --- 
generated by nonrelativistic distributions of matter.
In other words, it is imposed that Newtonian gravity should be recovered when the following 
criteria are met:
\begin{description}
\item[(i)]
The spacetime can be described by the line element
\begin{eqnarray}
ds^2 = (\eta_{\mu \nu} + h_{\mu \nu})dx^\mu dx^\nu,
\label{ds2N}
\end{eqnarray}
with $\eta_{\mu \nu } = {\rm diag}(-1,1,1,1)$, $|h_{\mu \nu}|\ll 1$, $|\partial_0h_{\mu\nu}|
%\approx 0$, 
\ll |\partial_j h_{\mu\nu}|$,
and only first-order terms in $h_{\mu\nu}$ (and their
spatial derivatives)
are considered
in the field equations;
\item[(ii)]
 Free-fall motion of a test particle is described by the nonrelativistic limit of the geodesic equation:
$$
\frac{d^2x^j}{dt^2} = -\partial^j \phi,
$$
where $\phi = -c^2 h_{00}/2$ (with $x^0 := ct$);
\item[(iii)] The {\it total} (physical)
stress-energy-momentum tensor curving the spacetime
satisfies $|T^{00}|\gg |T^{0j}|\gg|T^{jk}|$ in the coordinate basis associated to Eq.~(\ref{ds2N}).
\end{description}
When these conditions hold, $\phi$ appearing above plays the role of 
the Newtonian potential and must satisfy
\begin{eqnarray}
\partial^j\partial_j\phi= 4\pi G_N \rho_{m},
\label{NewtonEq}
\end{eqnarray}
where $\rho_m$ is the {\it observable} 
mass density of ordinary matter.

One might be tempted to apply the same criteria to obtain the Newtonian limit in our model. However, although
points (i) and (ii) can be consistently assumed, point (iii) is too restrictive
if $T^{ab}$ appearing in Eq.~(\ref{model1}) is the physical 
stress-energy-momentum tensor curving the spacetime.
%and in general will be violated.
This is easy to understand: if condition $|T^{00}|\gg |T^{0j}|\gg|T^{jk}|$ holds, then, from the trace
of Eq.~(\ref{modEqRab}), it follows that
$|R| \sim G_N|T^{00}|/(\alpha c^4)$, which then implies, according to our model,
$|\partial_\mu T^{\mu j}|\sim |\bar{\sigma}\partial^j T^{00}|$. In other 
words, spatial 
inhomogeneities in $T^{00}$ source the time-space and/or space-space
components of the stress-energy-momentum tensor, $T^{\mu j}$, so that condition 
$|T^{00}|\gg |T^{0j}|\gg|T^{jk}|$, even if initially true, may 
eventually be violated --- unless $\bar{\sigma}$ is negligible or we restrict attention to very 
special configurations. So, criterion (iii) should be replaced by:
\begin{description}
\item[(iii')] In the coordinate basis associated to Eq.~(\ref{ds2N}),
the {\it total} stress-energy-momentum tensor curving
the spacetime satisfies, to zeroth order in $h_{\mu\nu}$, the
equation obtained by combining  the trace of 
Eq.~(\ref{modEqRab}) and Eq.~(\ref{model1}) --- which is also the 
equation implied by the Bianchi identity applied to Eq.~(\ref{modEqGab}),
to zeroth order in $h_{\mu\nu}$:
\begin{eqnarray}
\partial_\mu T^{\mu \nu} =- \frac{\bar{\sigma}}{(1-4\bar{\sigma})}\partial^\nu T.
\label{model1Newtonian}
\end{eqnarray}
In particular, for a static distribution of perfect fluids
(as is customary to assume in obtaining the Newtonian regime), 
Eq.~(\ref{model1Newtonian}) can be integrated, leading to:
\begin{eqnarray}
(1-\bar{\sigma})P -\bar{\sigma}\rho =0,
%= 0,
\label{PrhoNewton}
\end{eqnarray}
where $P$ and $\rho$ are the {\it total} pressure and energy density
of the fluid distribution
(and the constant of integration has been 
fixed by the condition that $P = 0$ where $\rho = 0$ --- assuming such a region exists).
%The constant value above is determined by boundary conditions.
\end{description}

Some 
remarks are in order w.r.t.~point~(iii'). First, 
note that Eq.~(\ref{model1Newtonian}) shows that, for nonnegligible 
$\bar{\sigma}$, the behavior of the whole system ${\cal U}$ in the flat-spacetime 
{\it limit} (i.e., weak-gravity regime)
is quite different from its behavior  in exact
Minkowski spacetime (i.e.,  with gravity turned off), since, in the latter,
$\partial_\mu T^{\mu\nu} = 0$. 
%This reinforces the view that 
%$T^{ab}$ appearing in Eq.~(\ref{model1}) [and $T_{(j)}^{ab}$ appearing in 
%Eq.~(\ref{model1j}) is to be seen as the physical 
% stress-energy-momentum tensor of ${\cal U}$
Not surprisingly, 
Eq.~(\ref{model1Newtonian}) is at the heart of the criticism which says that
Rastall's proposal is inconsistent with flat-spacetime physics; it is taken to 
imply  
%either (a)~
that, conceding $T^{ab}$ to be the physical stress-energy-momentum tensor
--- which has been established previously to be perfectly plausible ---,
nonconservation would be ubiquitous 
in the flat-spacetime limit
%physics 
(for nonnegligible $\bar{\sigma}$), in sharp contradiction with observations~\cite{LH}. 
%or 
%(b)~that the {\it observable} stress-energy-momentum tensor would be
%(proportional to)
%${\cal T}^{ab} := T^{ab}+\bar{\sigma} g^{ab}T/(1-4\bar{\sigma})$ 
%--- which does satisfy
%$\nabla_a{\cal T}^{ab} = 0$ ---, instead
%of $T^{ab}$. 
%In summary, 
%Rastall's theory would be discarded by
%observations~\cite{LH}.
%or completely 
%equivalent to standard (conservative) GR~\cite{LH,Visser}.
However, it is clear that in the present Rastall-like model,
 Eq.~(\ref{model1Newtonian}) 
 is not supposed to hold for each and 
 every component ${\cal U}_j$ of our system;
 %${\cal U}
 %= \{{\cal U}_j\}$, 
 %but only for the {\it total} 
 %pressure and
 %energy density. 
 instead, the stress-energy-momentum tensor of each
 individual  constituent ${\cal U}_j$ should satisfy, to 
 zeroth order in $h_{\mu\nu}$ (and neglecting non-gravitational interactions
 among the constituents ${\cal U}_j$),
 \begin{eqnarray}
 \partial_\mu T_{(j)}^{\mu \nu} 
 =- \frac{\bar{\sigma}_j}{(1-4\bar{\sigma})}\partial^\nu T,
 \label{partialTmuj}
 \end{eqnarray}
 %(with $\bar{\sigma}_j:=\sigma_j/\alpha$). This equation can be rewritten
% or, equivalently, 
% %as 
% $$
% \partial_\mu\left[ T_{(j)}^{\mu \nu}+\frac{\bar{\sigma}_j}
% {(1-4\bar{\sigma})}\eta^{\mu \nu}T_{(j)} \right]
% =- \frac{\bar{\sigma}_j}{(1-4\bar{\sigma})}\partial^\nu T_{(\backslash{\!\!\!j})}.
% $$
% %where quantities with subscript
%%$\backslash{\!\!\!j}$ stand for their sum over $k\neq j$. 
%Therefore, following the
%rationale of Refs.~\cite{LH,Visser} adapted to our model, 
%${\cal T}^{\mu \nu}_{(j)}:=
%T_{(j)}^{\mu \nu}+\bar{\sigma}_j\eta^{\mu \nu}T_{(j)}/(1-4\bar{\sigma})$
%might pass for the physical stress-energy-momentum tensor of constituent 
%${\cal U}_j$ in the flat-spacetime limit.
 %$$
 %{\cal T}_{(j)}:=
%\frac{(1-4\bar{\sigma}_{\backslash{\!\!\!j}})}{(1-4\bar{\sigma})}T_{(j)}
% $$
 % $$
 %T^{\mu \nu}_{(j)}:={\cal T}^{\mu \nu}_{(j)}-
%\frac{\bar{\sigma}_j}
% {(1-4\bar{\sigma}_{\backslash{\!\!\!j}})}\eta^{\mu \nu}{\cal T}_{(j)} 
% $$
 which 
 for a perfect fluid with nonrelativistic spatial velocity $v^k\ll c$ implies
 \begin{eqnarray}
& & \frac{\partial}{\partial t}\left[\rho_{j}+\frac{\bar{\sigma}_j}{(1-4\bar{\sigma})}(\rho-3P)\right] = -
 \partial_k\left[(\rho_{j}+P_{j})v^k\right],
 \label{cons0}
 \\
 & & \partial^k\left[P_{j}-\frac{\bar{\sigma}_j}{(1-4\bar{\sigma})}(\rho-3P)\right] = -
 \frac{1}{c^2}\frac{\partial}{\partial t}\left[(\rho_{j}+P_{j})v^k\right]
 \label{consk}
 \end{eqnarray}
(the subscript $j$ indicates  quantities 
associated to ${\cal U}_j$). 
Therefore, 
for negligible values of $\bar{\sigma}_j$ --- for which
the second term in the left-hand side of Eqs.~(\ref{cons0}) and
(\ref{consk}) can be neglected ---, the condition 
$|P_{j}|\ll|\rho_{j}|$ can be consistently assumed in the 
nonrelativistic regime of ${\cal U}_j$ --- as observed for nonrelativistic ordinary matter. 
On the other hand, 
those constituents ${\cal U}_j$
with significative values of $\bar{\sigma}_j$
have their pressures $P_j$ sourced by spatial inhomogeneities on the spacetime scalar curvature $R$ (hence, by 
spatial inhomogeneities on the 
trace of the {\it total} stress-energy-momentum tensor), 
forcing them to  exert pressures which cannot be ignored 
($|P_j| \sim |\bar{\sigma}_j \rho|$).
In fact, considering the time-independent regime, 
%Eqs.~(\ref{PrhoNewton}) 
%and 
Eq.~(\ref{consk}) implies
\begin{eqnarray}
P_{j} = \frac{\bar{\sigma}_j\,\rho}{(1-\bar{\sigma})},
%\approx -\rho_{(j)}.
\label{PjNewton}
\end{eqnarray}
where the integration constant was fixed by imposing that $P_{j}=0$ 
where $\rho = 0$ (assuming there is such a region). 
Notice that it is the {\it total} 
energy density which appears 
in the r.h.s.\
of the equation above, not just $\rho_j$.
%.
%%This relation will  play an
%%important role in our discussion below.
%%Eq.~(\ref{PjNewton}) 
We see that any constituent ${\cal U}_j$ with nonnegligible value 
of $\bar{\sigma}_j$  would exhibit a behavior which we do not observe
for ordinary matter: pressure which cannot be ignored even in the 
nonrelativistic limit
%.
%We are led
%%, thus, to the following scenario:
%to conclude that 
%%ordinary matter which is observed to
%any such ${\cal U}_j$,
%%with nonnegligible $\bar{\sigma}_j$, 
%if it exists,
%should belong to
%the ``dark sector'' of the Universe's ingredients. 
%In summary: nonnegligible $\bar{\sigma}_j$ leads to a flat-space limiting
%behavior 
%of ${\cal U}_j$
%[Eq.~(\ref{PjNewton})] which is not observed for any {\it known} matter/interaction 
%field 
--- a fact which was used to dismiss Rastall's original
proposal in a time when dark matter and dark energy were topics
mainly dormant and
nonexistent, respectively. Notwithstanding, the Rastall-like model we consider here 
(with $T_{(j)}^{ab}$ in Eq.~(\ref{model1j}) being 
the physical stress-energy-momentum tensor of ${\cal U}_j$)  
can be well accommodated  in a Universe containing ``dark ingredients''
--- such as ours ---, provided ordinary matter  has 
negligible
$\bar{\sigma}_j$ and any ${\cal U}_j$ with a nonnegligible value of $\bar{\sigma}_j$ is hidden
as a ``dark ingredient.''

In order to proceed with the Newtonian-regime discussion, 
let us split the whole system ${\cal U}$ in 
%three 
two subsystems:
 ${\cal U}_{(c)} :=\{{\cal U}_j \,;\; \bar{\sigma}_j=0\}$ --- 
 the ``conservative'' 
 subsystem 
 %---,  
 %${\cal U}_{(ac)} :=\{{\cal U}_j \,;\; 0<|\bar{\sigma}_j|\ll 1\}$ --- 
 %the ``almost 
 %conservative'' subsystem 
 --- and 
 ${\cal U}_{(nc)} :=\{{\cal U}_j \,;\; 
 %|\bar{\sigma}_j|\text{ is significant}
 \bar{\sigma}_j \neq 0\}$
 --- the ``nonconservative'' subsystem ---, 
 with their own total
 pressures ($P_{(c)}$
 %$P_{(ac)}$, 
 and $P_{(nc)}$, respectively) 
 and energy densities
 ($\rho_{(c)}$
 %, $\rho_{(ac)}$, 
 and $\rho_{(nc)}$, respectively).
 Based on the discussion above, let us assume that the nonrelativistic ordinary-matter 
 density appearing in Eq.~(\ref{NewtonEq}) is given by
 $\rho_m = \rho_{(c)}/c^2$. Moreover, let $w_{0} :=
 P_{(nc)}/\rho_{(nc)}$  be the (effective) equation of state of
${\cal U}_{(nc)}$ in the nonrelativistic regime (i.e., when $P_{(c)} \approx 0$).
Then,
Eq.~(\ref{PjNewton}) summed over $j$
leads to
\begin{eqnarray}
\rho_{(nc)} \approx \frac{\bar{\sigma} \rho_{(c)}}{[w_{0}-\bar{\sigma}(w_{0}+1)]}.
%\approx -\rho_{(j)}.
\label{rhoncNewton}
\end{eqnarray}
This relation, valid in the nonrelativistic, weak-gravity regime,  shall play an important role
below.

Using  (i)--(iii')  
 %and Eqs.~(\ref{rhoNewton}) and (\ref{PNewton}) 
 to express the time-time component of Eq.~(\ref{modEqRab}) to first order in 
 $h_{\mu \nu}$, we obtain
 \begin{eqnarray}
\partial^j\partial_j \phi &=& \frac{4\pi G_N}{\alpha(1-4\bar{\sigma}) c^2}[(1-6\bar{\sigma})\rho
+3(1-2\bar{\sigma})P]
\nonumber \\
&=&\frac{4\pi G_N\,\rho}{\alpha(1-\bar{\sigma}) c^2}.
%-3\bar{\sigma}^2\rho]
%\nonumber \\
%&=&\frac{4\pi G_N }{\alpha(1-\bar{\sigma}) c^2}\,\rho\nonumber 
%\\
%&=&
%\frac{4\pi G_N}{\alpha (1-\bar{\sigma}_{(ac)}) c^2}\,\rho_{obs},
\label{modEqR00Newton}
\end{eqnarray}
The reader might be tempted to conclude, by comparing Eqs.~(\ref{NewtonEq})
and (\ref{modEqR00Newton}), that we should set 
$\alpha = 1/(1-\bar{\sigma}) = 1+\sigma$ in
order to recover the correct Newtonian regime.
%Unfortunately, things can be
%more complicated than that. The reason is that this conclusion would 
However, this would crucially depend
%on the identification $\rho/c^2 \equiv \rho_m$; i.e., 
on the assumption that
{\it all} constituents ${\cal U}_j$, including the ones with 
$\bar{\sigma}_j\neq 
0$, are  accounted for
in the {\it observed}
mass density in the Newtonian limit (i.e., when setting the experimental
value of $G_N$) ---  which we have already discussed above not to be
consistent with observations. Using Eq.~(\ref{rhoncNewton}) to relate
$\rho = \rho_{(c)}+\rho_{(nc)}$ with $\rho_m = \rho_{(c)}/c^2$, we  
obtain
%Combining all these considerations, Eq.~(\ref{modEqR00Newton}) can be
%rewritten in terms of $\rho_{(c)} = \rho_m c^2$:
\begin{eqnarray}
\partial^j\partial_j \phi \approx
\frac{4\pi G_N\,w_{0}\rho_m}{\alpha[w_{0}-\bar{\sigma}(w_{0}+1)]},
\label{modEqR00Newtonfinal}
\end{eqnarray}
from where we finally get the proper value of $\alpha$:
\begin{eqnarray}
\alpha = \frac{w_{0}}{[w_{0}-\bar{\sigma}(w_{0}+1)]} = 
1+ (1+w_{0}^{-1})\sigma.
\label{alpha}
\end{eqnarray}
Substituting this result into Eq.~(\ref{modEqGen2}), we finally 
obtain the modified
Einstein's equations for our Rastall-like model 
which recover Newtonian gravity in the proper limit:
\begin{eqnarray}
G^{ab}+\sigma \left[(1+w_0^{-1})R^{ab}+\frac{1}{2}(1-w_0^{-1})g^{ab}R\right]
= \frac{8\pi G_N}{c^4}T^{ab},
\label{modEqGR}
\end{eqnarray}
or, equivalently [Eq.~(\ref{modEqGab})], 
\begin{eqnarray}
G^{ab}
= \frac{8\pi G_N}{[1+ (1+w_{0}^{-1})\sigma]c^4}\left[ T^{ab}
+\frac{\sigma g^{ab}T}{1-(3-w_{0}^{-1})\sigma}
\right].
\label{modEqG}
\end{eqnarray}
The latter equation is very convenient in some applications 
since it maps our Rastall-like model to standard GR with
the {\it effective} stress-energy-momentum tensor
\begin{eqnarray}
T_\text{\it eff}^{ab}:=
\frac{1}{[1+ (1+w_{0}^{-1})\sigma]}\left[ T^{ab}
+\frac{\sigma g^{ab}T}{1-(3-w_{0}^{-1})\sigma}
\right]
%\nonumber \\
%&=&
%\bar{T}^{ab}+\frac{g^{ab}T/4}{(1-4\sigma)}
%\\
%R^{ab}&=&\frac{8\pi G}{c^4}\left[ 
%\bar{T}^{ab}-\frac{g^{ab}T/4}{(1-4\sigma)}
%\right]
\label{Teff}
\end{eqnarray}
%This should not be confused, though, with the claim that
%these Rastall-like models and GR are physically equivalent~\cite{LH,Visser}. 
%{\it Any} metric theory of gravity
%can be put in the form $G^{ab} = 8\pi G_N {\cal T}^{ab}/c^4$, with 
%${\cal T}^{ab}$ some tensor satisfying $\nabla_a{\cal T}^{ab}=0$ --- 
%simply solve whatever the gravity field equations are, calculate
%$G^{ab}$ from the solution, and then extract ${\cal T}^{ab}$ out of 
%$ {\cal T}^{ab} = c^4 G^{ab}/(8\pi G_N)$.
%This trivial fact does {\it not} allow one to interpret ${\cal T}^{ab}$ as
%the {\it physical} stress-energy-momentum tensor --- a quantity which must
%make
%sense (and, therefore, be defined) even in the complete absence of gravity.
%Arguing that whatever ${\cal T}^{ab}$ satisfying 
%$G^{ab} = 8\pi G_N {\cal T}^{ab}/c^4$
%can be seen as the physical stress-energy-momentum tensor would reduce GR to 
%a mere tautology.
%Returning to the useful form given in Eq.~(\ref{Teff}),
--- stressing, again, that this does {\it not} mean that our model is
equivalent to GR  for the given constituents $\{{\cal U}_j\}$
(see Subsec.~\ref{subsec:NotEquiv}).
In case  $T^{ab}$ 
%and $T^{ab}_{(nc)}$ 
assumes the perfect-fluid form with a common 4-velocity field for all constituents 
${\cal U}_j$
(as in the next sections),
%with
%the same 4-velocity field (as  in the next sections), then 
the effective energy density
and pressure are given by
\begin{eqnarray}
\rho_\text{\it eff} &=& \frac{(\alpha-3\sigma)\rho-3\sigma P}{\alpha(\alpha-4\sigma)},
\label{rhoeff}
\\
P_\text{\it eff} &=& \frac{(\alpha-\sigma)P-\sigma \rho}{\alpha(\alpha-4\sigma)},
\label{Peff}
\end{eqnarray}
with $\alpha$ given by Eq.~(\ref{alpha}).

Before exploring strong gravity consequences of this simple model,
%given by Eq.~(\ref{model1}), 
let us briefly  discuss the parameter $w_0$.

\section{Nonrelativistic equation of state  for ${\cal U}_{(nc)}$}
\label{sec:NRw}

In the previous section, we have seen that if $\sigma$ is nonnegligible, then
the pressure $P_{(nc)}$ associated to the nonconservative constituents ${\cal U}_{(nc)}$
cannot, in general, be neglected as a source of spacetime curvature even in the socalled Newtonian regime
(when $P_{(c)}\approx 0$). 
%Assuming  energy densities are positive,
%Eqs.~(\ref{PrhoNewton}) and (\ref{alpha}) imply
%\begin{eqnarray}
%w_0/\sigma >0 \;\;\;\text{or}
%\;\;\; w_0/\sigma <-2.
%\label{alphabound}
%\end{eqnarray}
(An exception to this rule appears in the special case of
homogeneous cosmology, as we shall see in Sec.~\ref{sec:modcosm}.)

There are different approaches we can take regarding the value of $w_0$. We can consider it to be fixed by the nature
of ${\cal U}_{(nc)}$ --- in which case, momentum densities and energy currents arise, in the course of nonrelativistic evolution
[governed by Eqs.~(\ref{cons0}) and (\ref{consk})], so that $P_{(nc)}/\rho_{(nc)}= w_0$ is ensured. Or, alternatively, 
we can consider that
$\rho_{(nc)}$ and $P_{(nc)}$ can be independently sourced 
according to Eq.~(\ref{model1}). In this latter scenario,
$P_{(nc)}/\rho_{(nc)}$ may assume different values depending on the
details of the evolution. A particularly interesting
situation is the one in which momentum densities and energy currents, in
the r.h.s.\ of Eqs.~(\ref{cons0}) and (\ref{consk}), can 
be neglected in comparison to the second term in the l.h.s.\ of the corresponding
equations (for some particular constituent ${\cal U}_j\in{\cal U}_{(nc)}$). Then,
using an initially and distantly diluted 
($\rho\approx P\approx 0$)
configuration as initial and boundary conditions, 
respectively,  to evolve the system 
according to Eqs.~(\ref{cons0}) and (\ref{consk}), we get the implicit relations
\begin{eqnarray}
\rho_j &\approx& -  \frac{\bar{\sigma}_j}{(1-4\bar{\sigma})}(\rho-3P),
\label{rhojNIBC}
\\
P_j &\approx&  \frac{\bar{\sigma}_j}{(1-4\bar{\sigma})}(\rho-3P);
\label{PjNIBC}
\end{eqnarray}
i.e., regardless the value of $\bar{\sigma}_j$, this constituent ${\cal U}_j$
would have an effective equation of state given by $P_j \approx -\rho_j$ in this regime. It is 
an interesting (if not intriguing) coincidence 
that such a natural condition for the nonrelativistic
evolution (namely, negligible momentum densities and 
energy currents) would force a constituent with a significative 
value of $\bar{\sigma}_j$ to exhibit the same equation of state 
$P_j/\rho_j \approx -1$ which seems to be needed to account for the large-scale
behavior of our Universe. In case all nonconservative constituents 
 behave like this, we would have $w_0 \approx -1$; 
but, in general, $w_0$ will be the average of the equations of state of 
${\cal U}_j\in {\cal U}_{(nc)}$.

\section{Residual cosmological constant from $P = -\rho$ equation of state}
\label{sec:cc}

Before jumping to the analysis of how Eqs.~(\ref{model1}) 
and (\ref{model1j}),
with the correct Newtonian limit, affects standard 
scenarios of gravitation, such as the Schwarzschild 
and the Friedmann-Lema\^itre-Robertson-Walker (FLRW) solutions, let us consider
the effects of a given nonconservative constituent ${\cal U}_k\in {\cal U}_{(nc)}$ 
with equation of state 
$P_k = -\rho_k$. The motivation for this is twofold. On one hand, it 
has been shown, in the previous section, that
 a nonconservative constituent  ${\cal U}_k$
with a significative value of
$\sigma_k$, for which 
momentum densities and energy currents can be neglected in the Newtonian regime,
has energy density and pressure satisfying
$ P_k \approx -\rho_k$ 
[Eqs.~(\ref{rhojNIBC}) and (\ref{PjNIBC})]. 
On the other hand,
in standard GR,
a ``dark'' constituent with such an equation of state --- the cosmological constant
--- is needed to explain the observed accelerating cosmic expansion. Therefore, it 
is only natural to explore the effects of such a fixed equation of state in the present 
model.

The stress-energy-momentum tensor 
of a perfect fluid with $P_k = -\rho_k$ is  $T_{(k)}^{ab} = 
-\rho_k g^{ab}$, 
which plugged into Eq.~(\ref{model1j}) (with $f_{(k)}^a = 0$, for simplicity) leads
to $\rho_k = -c^4 \sigma_k 
R /(8\pi G_N)+C_k$, where $C_k$ is a mere integration constant. 
Substituting this result into Eq.~(\ref{modEqGen2}), we get
\begin{eqnarray}
G^{ab}+\Lambda_kg^{ab}+\bar{\sigma}_{\backslash{\!\!\!k}}\, g^{ab}R
= \frac{8\pi G_N}{\alpha c^4}T_{(\backslash{\!\!\!k})}^{ab},
\label{EEqCC}
\end{eqnarray}
with $\Lambda_k :=8\pi G_N C_k/(\alpha c^4)$.
%and where quantities with subscript
%$\slash{\!\!\!k}$ stand for their sum over $j\neq k$. 
In words, Eq.~(\ref{EEqCC}) shows that 
a nonconservative constituent with
equation of state $P_k = -\rho_k$ would fall almost into oblivion, regardless its
energy density $\rho_k$, except, possibly, 
for a constant term which leads to a cosmological-constant-like contribution.
Note that, although 
similar, this is different than  what happens in standard GR: in the 
latter, $P_k = -\rho_k$
implies that the {\it whole}
energy density $\rho_k$ is kept constant and contributes to the
cosmological-constant term $\Lambda = 8\pi G_N\rho_k/c^4$. This difference 
may alleviate the problem of the naturalness of the value of the 
cosmological
constant $\Lambda$, since in the present model most of $\rho_k$ might simply 
drop out from the dynamical equations, leaving only a residual contribution
$\Lambda_k=8\pi G_N C_k/(\alpha c^4)$. (Note, in particular, that if such
${\cal U}_k$ is the only nonconservative constituent, then $\alpha = 1$,
$\bar{\sigma}_{\backslash{\!\!\!k}}=0$, and $T_{(\backslash{\!\!\!k})}^{ab} = T_{(c)}^{ab}$.)
%We shall elaborate more on this in  
%Sec.~\ref{sec:modcosm}.

\section{Modified Schwarzschild solution}
\label{sec:modSch}

As we  pointed out earlier, our modified Einstein's equations possess the 
same vacuum
solutions as standard GR. Therefore, the line element outside a static,
spherically-symmetric body is given by the usual exterior Schwarzschild solution:
\begin{eqnarray}
ds^2 = -\left(1-\frac{2G_N M_g}{c^2 r}\right) c^2 dt^2+
\left(1-\frac{2G_N M_g}{c^2 r}\right)^{\!\!-1}dr^2 +r^2d\Omega^2,
\label{extSchw}
\end{eqnarray}
where $d\Omega^2 = d\theta^2+(\sin\theta)^2 d\varphi^2$ is the line element of the
unit 2-sphere --- $\{(t,r,\theta,\varphi)\}$ being 
the usual Schwarzschild coordinate
system --- and
$M_g$ is an integration constant (the ``gravitational mass'') 
which in the Newtonian limit is given by the
mass of the spherical body in the form of {\it ordinary} 
%(i.e., non-dark) 
matter (see Subsec.~\ref{subsec:newton}).
%(i.e., which does not
%exchange 4-momentum with the spacetime; 
%(Note from Eq.~(\ref{rhoNewton}) and
%the discussion
%below Eq.~(\ref{PjNewtonFinal}) that 
%the true mass of the spherical body in the Newtonian
%regime would be 
%$M_T = (1-\sigma)M$.)

In the interior region, spherical symmetry and staticity allow  
the line element to be put in the form
\begin{eqnarray}
ds^2 = -e^{2\phi(r)/c^2}c^2 dt^2+g(r) dr^2+r^2d\Omega^2,
\label{intSchw}
\end{eqnarray}
where  $\phi(r)$ and $g(r)$ must be determined from Eq.~(\ref{modEqGR}).
%and Eq.~(\ref{model1j}) applied to each ${\cal U}_j$. 
Mapping this problem to the
one of standard GR with effective stress-energy-momentum tensor given by
Eqs.~(\ref{Teff})-(\ref{Peff}), we have:
\begin{eqnarray}
g(r) &=& \left(1-\frac{2G_N }{c^2r}m_\text{\it eff}(r)\right)^{\!\!-1},
\label{gr}
\\
\frac{d}{dr}\phi(r) &=&  g(r)\left(\frac{G_N m_\text{\it eff}(r)}{r^2}+\frac
{4\pi 
G_N\, r}{c^2} 
P_\text{\it eff}(r) \right),
%\nonumber \\ 
%&=&
%g(r)\left(\frac{G_N m_\text{\it eff}(r)}
%{r^2}+\frac{4\pi G_N\, r}{(1-4\sigma)c^2} 
% [(1-\sigma)P(r)-\sigma \rho(r)] \right),
\label{dphidr}
\\
c^2\frac{d}{dr}P_\text{\it eff}(r)&=& -[\rho_\text{\it eff}(r)+
P_\text{\it eff}(r)] \frac{d}{dr}\phi(r),
\label{conseff}
\end{eqnarray}
where
\begin{eqnarray}
m_\text{\it eff}(r) &=& 4\pi \int_0^r dr'r'^2\rho_\text{\it eff}(r')/c^2.
%\nonumber \\
%& =& \frac{4\pi}{(1-4\sigma)}\int_0^r dr' r'^2\left[(1-3\sigma)\rho(r') -3\sigma 
%P(r')
%\right]/c^2.
\label{meff}
\end{eqnarray}
Note that $M_g$ appearing in Eq.~(\ref{extSchw}) is set by $M_g = 
m_\text{\it eff}(R)$, where $r = R$ determines the surface of the central object.

The reader, noticing that $\sigma$ appears nowhere explicitly 
in Eqs.~(\ref{gr})-(\ref{meff}), 
might 
%be tempted to 
conclude that
the effect of this parameter gets ``screened''
%as far as the 
from the spacetime geometry,
%is concerned,
so that 
nothing 
interesting
%new 
will arise from solving these equations 
in comparison to standard GR --- 
apparently siding with the view of equivalence 
between Rastall-like models and GR.
%(once we 
%make
%the 
%``simple concession'' that the gravitational mass would be given by 
%$m_\text{\it eff}$ instead
%of integrating $\rho$ alone). 
However, in line with our discussion in Subsec.~\ref{subsec:NotEquiv},
we should recall, from standard GR,
that
Eqs.~(\ref{gr})-(\ref{meff}) constitute an {\it underdetermined} system of 
integro-differential equations for the functions $g$, $\phi$, $\rho_\text{\it eff}$,
$P_\text{\it eff}$, and $m_\text{\it eff}$; additional information must be 
provided
in order to determine their solutions. 
Such additional information is usually
provided in the form of an equation of state, relating pressure and 
energy density, or
some other similar condition (e.g., uniformity of the energy 
density distribution).
And this is how $\sigma$ comes into play: this additional information is
more naturally given in terms of $\rho$ and $P$ --- actually, in terms
of $\rho_j$ and  $P_j$ 
for each constituent ${\cal U}_j$ --- rather than in terms of 
$\rho_\text{\it eff}$ and $P_\text{\it eff}$. The net effect of this is that 
the additional
condition provided to the system of Eqs.~(\ref{gr})-(\ref{meff}) will 
almost inevitably involve 
$\sigma$ when expressed in terms of $\rho_\text{\it eff}$
and $P_\text{\it eff}$.
%Another way of seeing this is noticing that
In fact, 
Eq.~(\ref{conseff}) is actually the combination
of as many  equations as there are constituents (Eq.~(\ref{model1j}) for each 
${\cal U}_j$); 
in particular, in terms of ${\cal U}_{(c)}$ and 
${\cal U}_{(nc)}$,
we have
\begin{eqnarray}
c^2\frac{dP_{(c)}}{dr} &=& -(\rho_{(c)}+P_{(c)}) \frac{d\phi}{dr},
\label{consc}
\\
%c^2\frac{dP_{(nc)}}{dr} &=& -(\rho_{(nc)}+P_{(nc)}) \frac{d\phi}{dr}
%+\frac{\bar{\sigma} c^2}{(1-4\bar{\sigma})}\frac{d}{dr}\left(\rho-3P\right),
c^2\frac{d}{dr}
\left[
\frac{(1-\bar{\sigma})P_{(nc)}-\bar{\sigma} \rho_{(nc)}}{(1-4\bar{\sigma})}
\right]
&=&-(\rho_{(nc)}+P_{(nc)}) \frac{d\phi}{dr}
+\frac{\bar{\sigma} c^2}{(1-4\bar{\sigma})}
\frac{d}{dr}\left(\rho_{(c)}-3P_{(c)}\right),
\label{consnc}
\end{eqnarray}
where the latter equation comes from substituting  the trace of
Eq.~(\ref{modEqRab}) into Eq.~(\ref{model1j}) for ${\cal U}_{(nc)}$ and 
rearranging some terms 
[or, equivalently, from combining Eqs.~(\ref{conseff}) and (\ref{consc}), recalling Eqs.~(\ref{rhoeff}) and (\ref{Peff})].

In order to concretely 
illustrate the possible effects of $\sigma$ in this static, 
spherically-symmetric scenario, let us consider our central body to be 
constituted by
``ordinary matter'' (making up ${\cal U}_{(c)}$) and some ``dark constituent''
which exchange 4-momentum with the spacetime (making up ${\cal U}_{(nc)}$).
For ${\cal U}_{(c)}$ we impose $\rho_{(c)} =\rho_0 = \text{ constant}$ for
$r<R$ (the radius of the central object), 
whereas for 
${\cal U}_{(nc)}$ we consider two scenarios: (i) one with barotropic 
equation of state $P_{(nc)} = w_0\rho_{(nc)}$, with $w_0$ constant,
%--- just for simplicity and
%because this relation is forced on ${\cal U}_{(nc)}$ in the Newtonian 
%regime
%(see Eqs.~(\ref{rhojNewton}) 
%and (\ref{PjNewton}) and the discussion following them).
%\begin{eqnarray}
%c^2\frac{d}{dr} \left[
%P-\sigma (\rho+P)\right]&=& -(1-4\sigma)(\rho+P) \frac{d\phi}{dr},
%\label{consNC}
%\end{eqnarray}
and (ii) one in which $P_{(nc)} = -\rho_{(nc)}$ in the region 
where
$P_{(c)}\approx 0$
[motivated by the nonrelativistic analysis which led to Eqs.~(\ref{rhojNIBC}) and 
(\ref{PjNIBC})]
but with $\rho_{(nc)}$~constant --- hence, given by $\rho_{(nc)} = 
-\sigma\rho_{0}$ (see Eqs.~(\ref{rhoncNewton}) 
and (\ref{alpha}) with $w_0 = -1$) ---
throughout the central object.

\vskip 0.3cm
\noindent
%$\bullet$ 
{\bf (i)}~{\bf{\underline{$P_{(nc)} = w_0 \rho_{(nc)}$}:}}
\vskip 0.3cm
\noindent
%Also, assuming that ${\cal U}_{(c)}$ and ${\cal U}_{(nc)}$
%do not directly interact with each other, Eq.~(\ref{model1j}), combined
%with 
Substituting $\rho_{(c)} = \rho_0$ and $P_{(nc)} = w_0 \rho_{(nc)}$ into 
Eqs.~(\ref{consc}) and (\ref{consnc}), combining the results, and using Eq.~(\ref{alpha}), we obtain
\begin{eqnarray}
\frac{d\rho_{(nc)}}{dr}
-\left[1-\sigma(3-w_0^{-1})\right]\left(1+w_0^{-1}\right)
\frac{\rho_{(nc)}}{(\rho_0+P_{(c)})} \frac{dP_{(c)}}{dr}
=
%\ln\left|1+\frac{P_{(c)}}{\rho_0}\right|
-3\sigma w_0^{-1} 
\frac{dP_{(c)}}{dr},
\label{conscnc}
\end{eqnarray}
which can be integrated to give~\footnote{
The particular case in which $1+\sigma(w_0+1)(w_0^{-1}-3)=0$ leads to
$$
P_{(nc)}=w_0\rho_{(nc)} = \sigma (\rho_0+P_{(c)})\left[1-3\ln\left|1+\frac{P_{(c)}}{\rho_0}\right|\right],
$$
as can be easily verified by direct integration of Eq.~(\ref{conscnc}) or by a limiting process 
in
Eq.~(\ref{rhoncPcw0}).}
\begin{eqnarray}
P_{(nc)} = w_0\rho_{(nc)} &=&\sigma (\rho_0+P_{(c)})
\left\{
\frac{3 w_0}{1+\sigma (w_0+1)(w_0^{-1}-3)}
\right.
\nonumber \\ 
& &\;\;\;\;\;\;\;\;\;\;\;\;\;\;\;\;\;\;\;\;\;\;
\left.+
\left[1
-
\frac{3 w_0}{1+\sigma(w_0+1)(w_0^{-1}-3)}
\right] \left(1+\frac{P_{(c)}}{\rho_0}\right)^{w_0^{-1}+\sigma(1+w_0^{-1})(w_0^{-1}-3)}
\right\},
\label{rhoncPcw0}
\end{eqnarray}
where we have already imposed the boundary condition $\left.\rho_{(nc)}\right|_R = \sigma w_0^{-1} \rho_0$
at the surface of the central object (where $P_{(c)} = 0$). This boundary condition can be inferred either from 
Eq.~(\ref{PrhoNewton}) or directly through Eq.~(\ref{consnc}), which imposes that any ``jump'' $\Delta \rho_{(c)}$
in $\rho_{(c)}$ should be accompanied by ``jumps'' $\Delta \rho_{(nc)}$ and $\Delta P_{(nc)}$ in $\rho_{(nc)}$
and $P_{(nc)}$, respectively, satisfying $(1-\bar{\sigma})\Delta P_{(nc)}-\bar{\sigma} \Delta \rho_{(nc)} = \bar{\sigma}
\Delta \rho_{(c)}$ [recalling that $P_{(c)}$ is continuous as a result of Eq.~(\ref{consc})].

\begin{figure}
\includegraphics[scale=0.55]{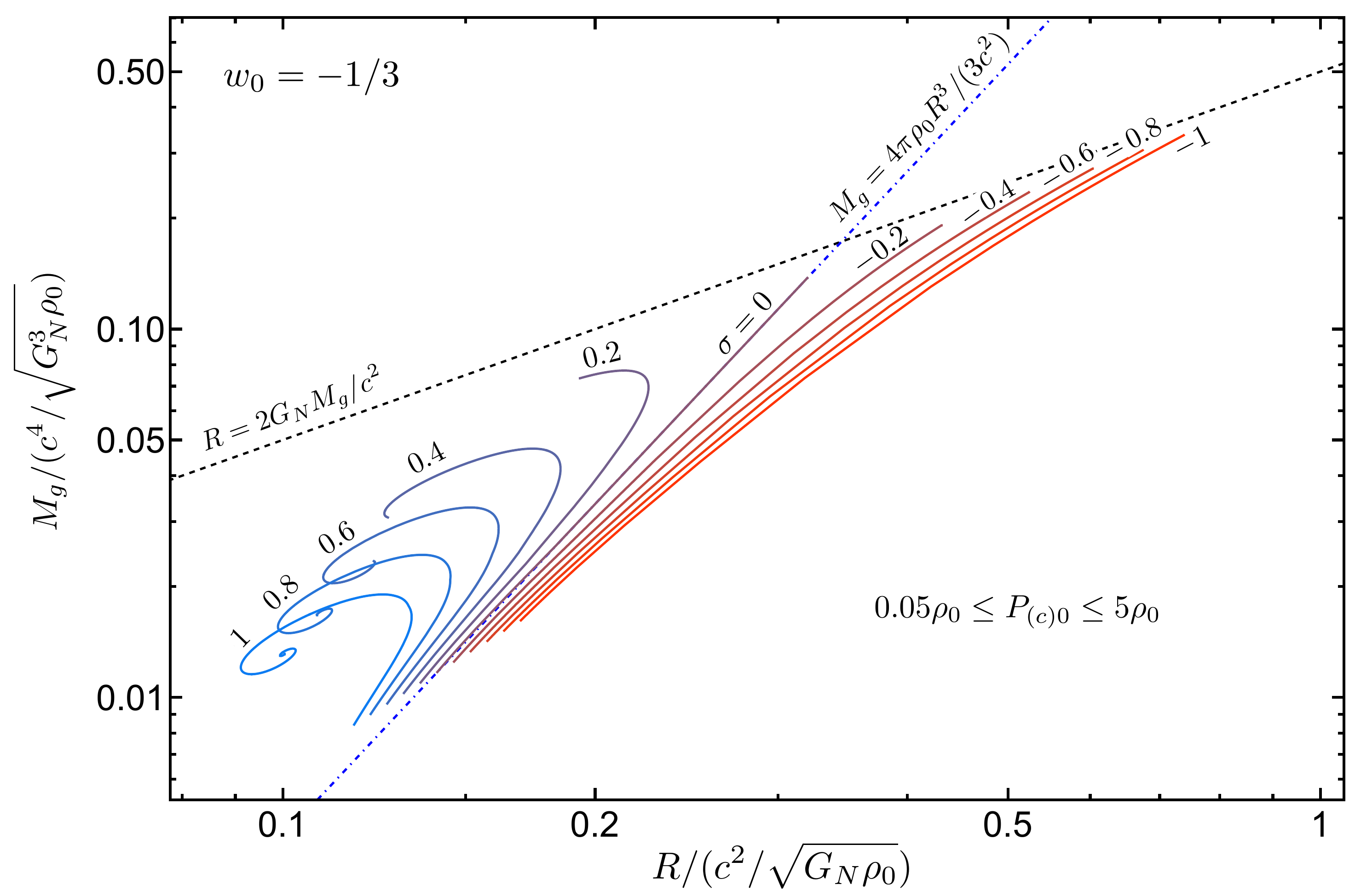}
\caption{Phase diagram showing the effective 
gravitational mass ($M_g$) and radius ($R$) of a spherical 
object, with uniform ordinary-matter density $\rho_0$, 
in hydrostatic equilibrium, for $w_0 = -1/3$. 
Each solid curve represents  equilibrium 
configurations
for values of 
ordinary-matter 
central pressure in the range $0.05 \rho_0< P_{(c)0}<5\rho_0$, 
for a given value
of $\sigma$ (indicated next to the corresponding curve). As expected, 
no equilibrium configuration is possible beyond the Schwarzschild limit
($R = 2G_N M_g/c^2$;  dashed line). The dot-dashed line gives the
ordinary mass contained in an object with radius $R$ and uniform energy density
$\rho_0$ --- which coincides with the gravitational mass an object with the same
density and radius would have (if possible) in standard GR.
}
\label{fig:w0m1by3diag}
\end{figure}

\begin{figure}
\includegraphics[scale=0.55]{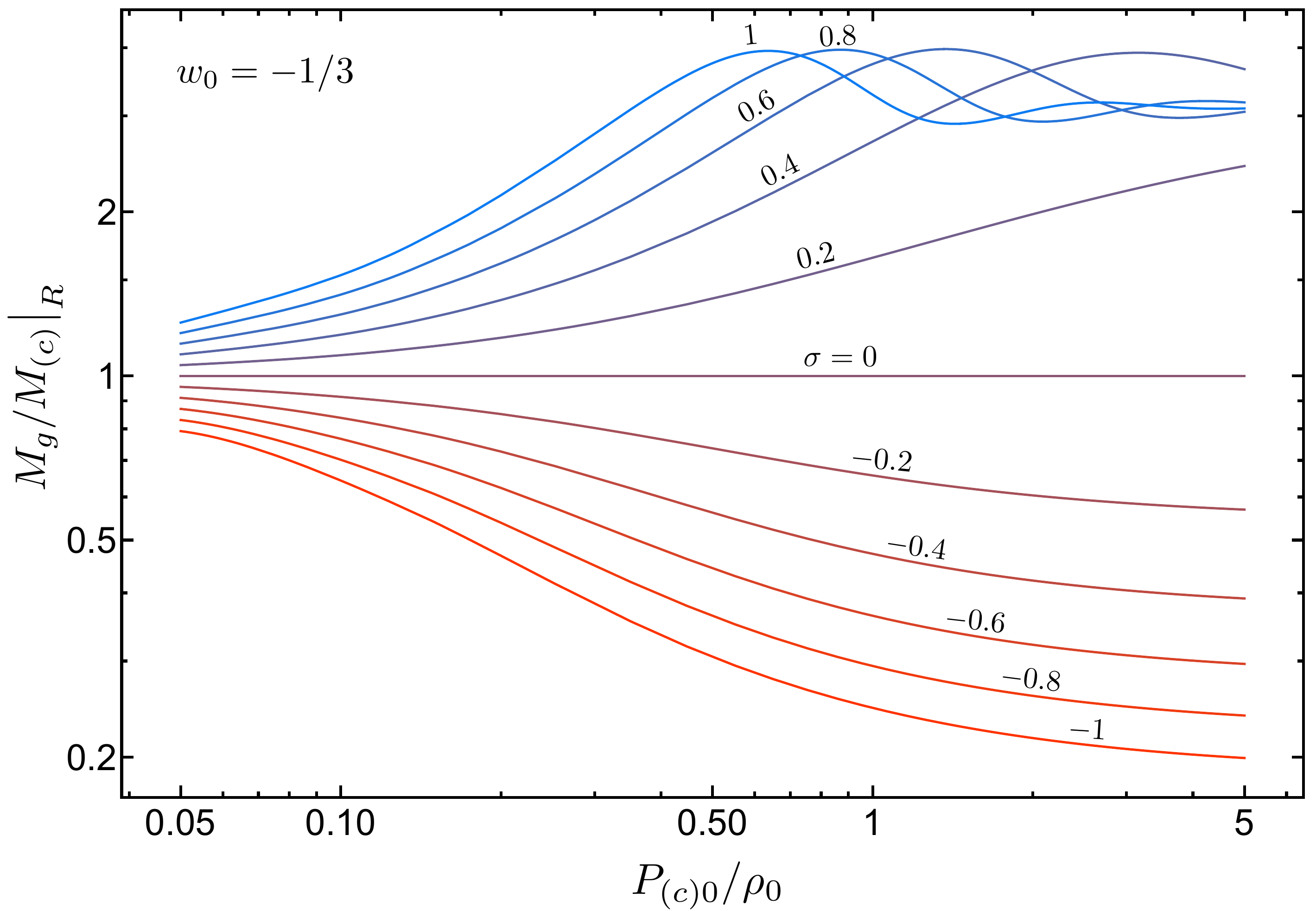}
\caption{Ratio between the effective gravitational mass $M_g$ of a central body with uniform 
ordinary-matter density
$\rho_0$, in case $w_0 = -1/3$, and its ordinary mass $M_{(c)} = 4\pi\rho_0 R^3/(3c^2)$ --- the latter also being
the gravitational mass
a body with same $\rho_0$ and same radius $R$ would have (if possible)
in standard GR. 
Each curve represents a given value of $\sigma$ (indicated next to the corresponding curve).}
\label{fig:w0m1by3MMR}
\end{figure}

\begin{figure}
\includegraphics[scale=0.55]{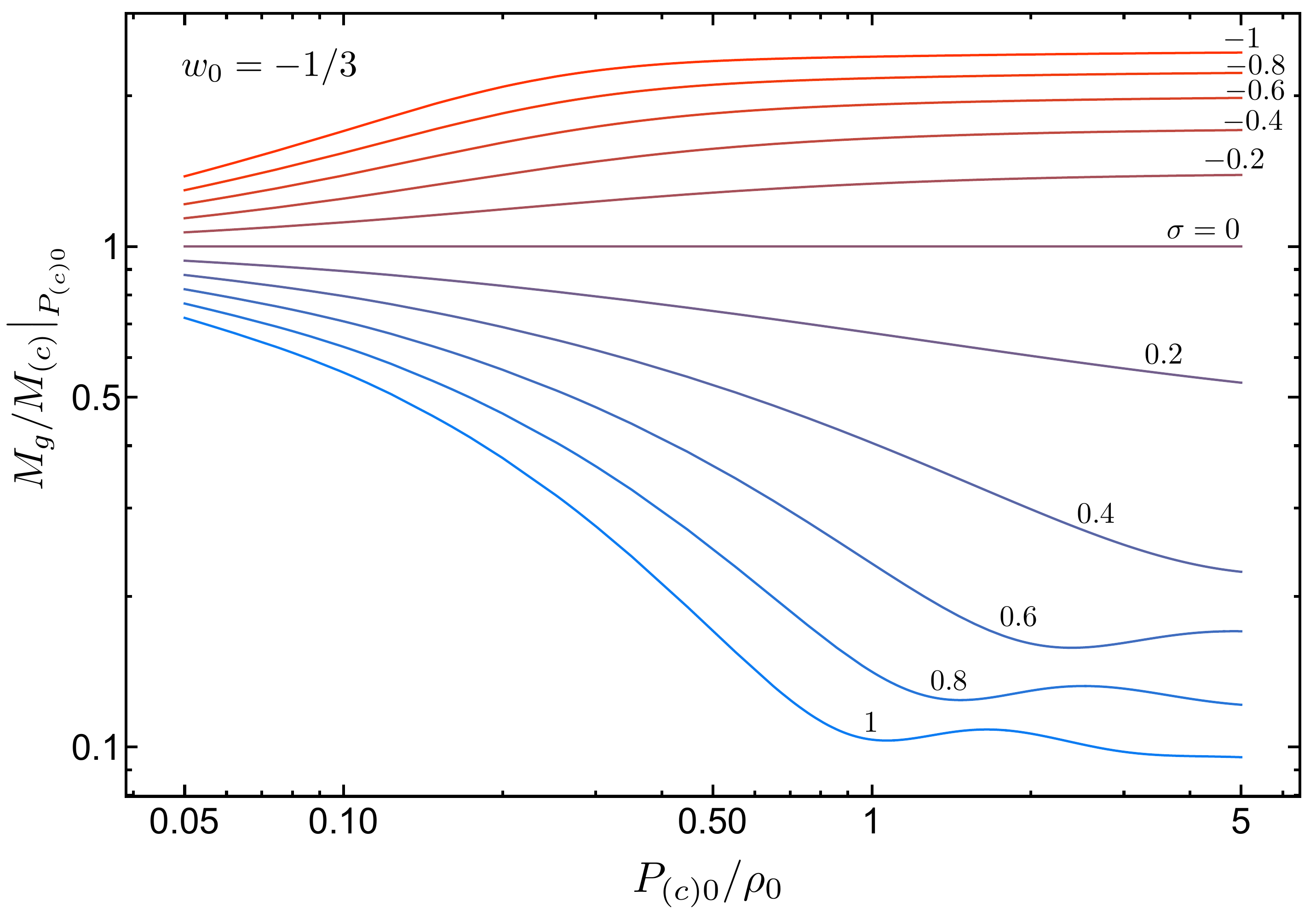}
\caption{Ratio between the effective gravitational mass $M_g$ of a central body with uniform 
ordinary-matter density
$\rho_0$, in case $w_0 = -1/3$, and the ordinary mass $M_{(c)} = 4\pi\rho_0 R^3/(3c^2)$ 
a body with same $\rho_0$ and same central pressure $P_{(c)0}$ would have
in standard GR. 
Each curve represents a given value of $\sigma$ (indicated next to the corresponding curve).}
\label{fig:w0m1by3MMP0}
\end{figure}

\begin{figure}
\includegraphics[scale=0.55]{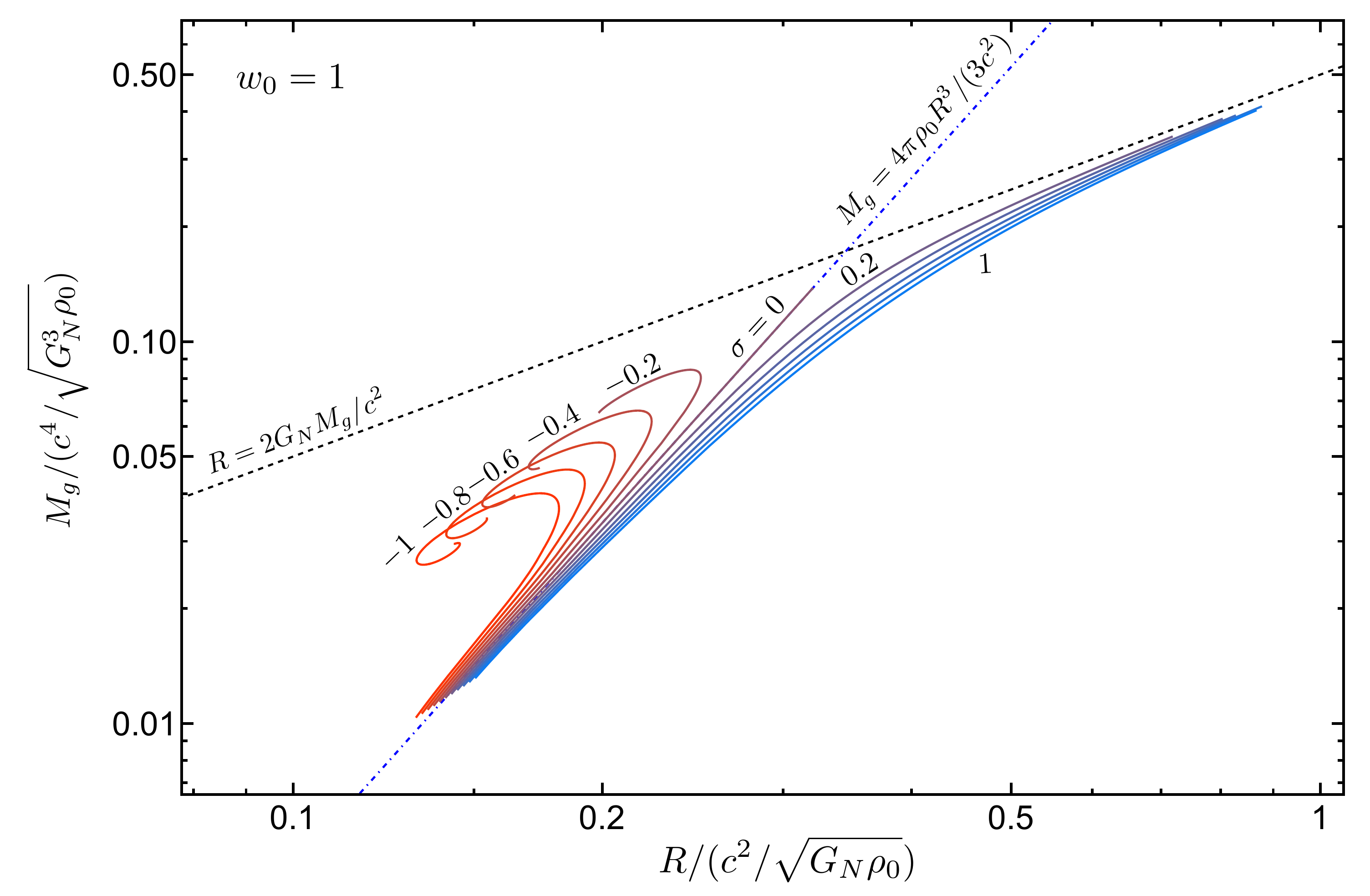}
\caption{Phase diagram showing the effective 
gravitational mass ($M_g$) and radius ($R$) of a spherical 
object, with uniform ordinary-matter density $\rho_0$, 
in hydrostatic equilibrium, for $w_0 = 1$. 
Each solid curve represents  equilibrium 
configurations
for varying  
ordinary-matter 
central pressures ($P_{(c)0}$), 
for a given value
of $\sigma$ (indicated next to the corresponding curve, when possible; 
the curves in between the ones with $\sigma = 0.2$ and $\sigma = 1$
correspond to $\sigma = 0.4$, $0.6$, and $0.8$, in this order). As expected, 
no equilibrium configuration is possible beyond the Schwarzschild limit
($R = 2G_N M_g/c^2$;  dashed line). The dot-dashed line gives the
ordinary mass contained in an object with radius $R$ and uniform energy density
$\rho_0$ --- which coincides with the gravitational mass an object with the same
density and radius would have (if possible) in standard GR.}
\label{fig:w01diag}
\end{figure}

\begin{figure}
\includegraphics[scale=0.55]{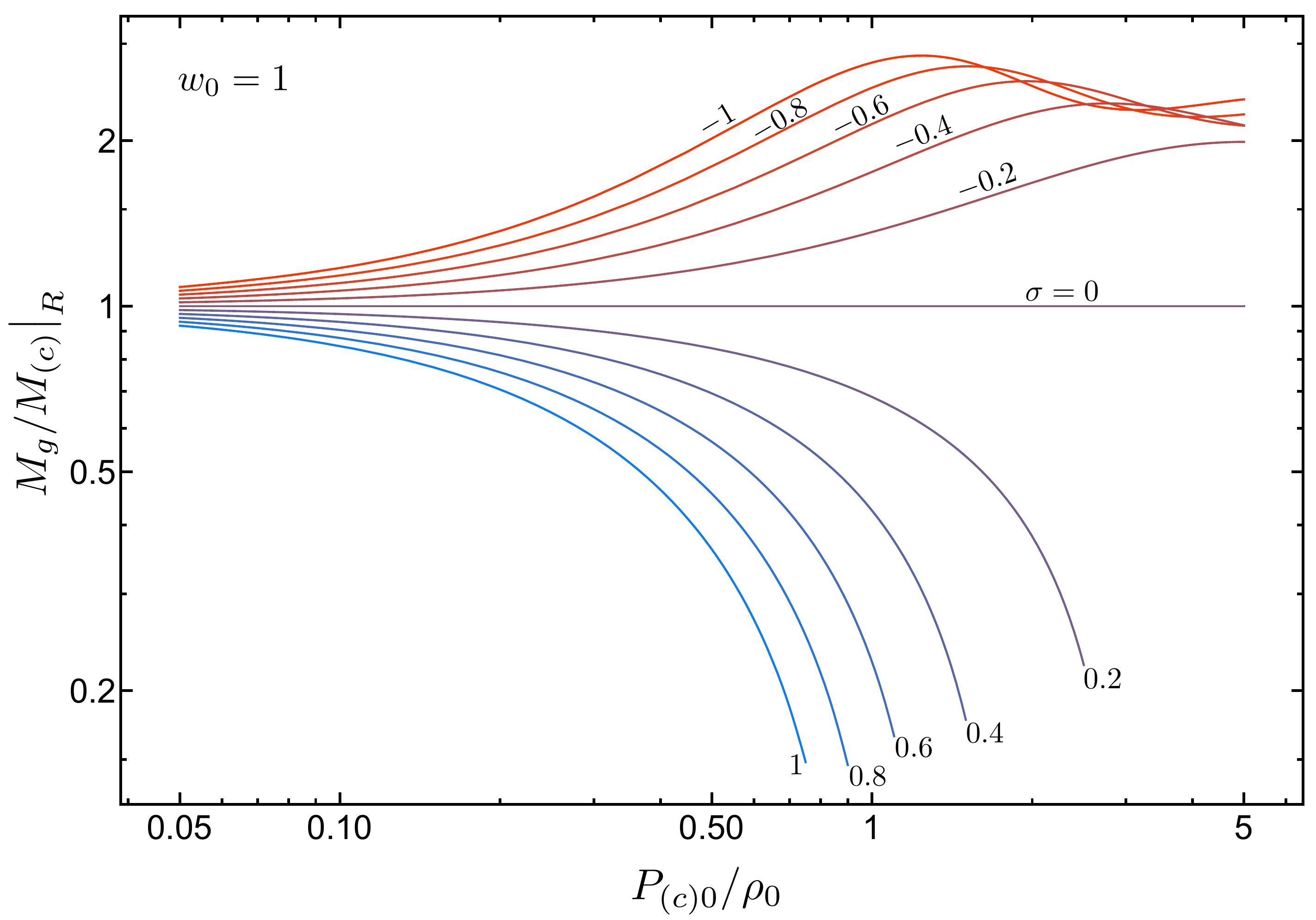}
\caption{Ratio between the effective gravitational mass $M_g$ of a central body with uniform 
ordinary-matter density
$\rho_0$, in case $w_0 = 1$, and its ordinary mass $M_{(c)} = 4\pi\rho_0 R^3/(3c^2)$ --- the latter also being
the gravitational mass
a body with same $\rho_0$ and same radius $R$ would have (if possible)
in standard GR. 
Each curve represents a given value of $\sigma$ (indicated next to the corresponding curve).}
\label{fig:w01MMR}
\end{figure}

\begin{figure}
\includegraphics[scale=0.55]{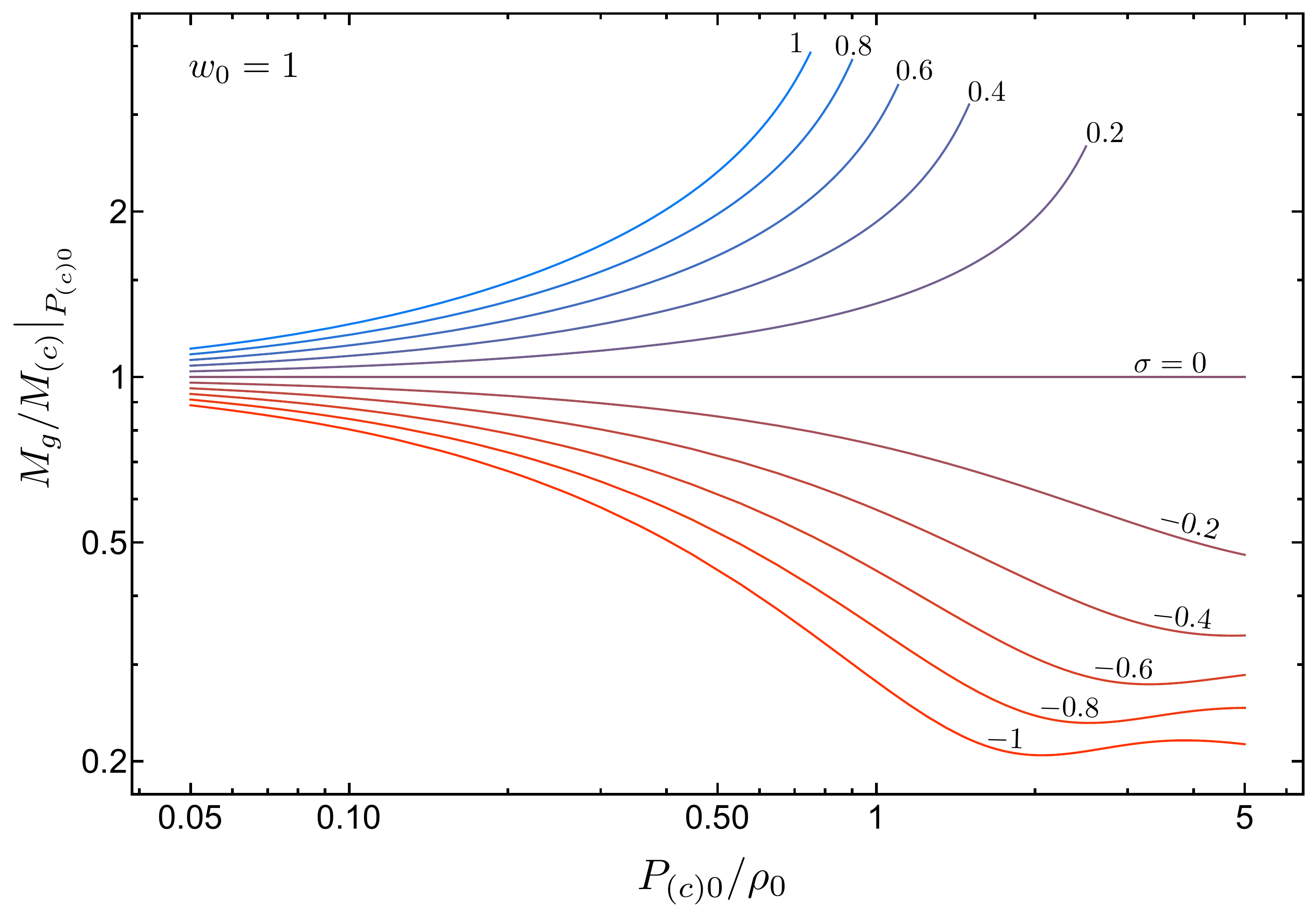}
\caption{Ratio between the effective gravitational mass $M_g$ of a central body with uniform 
ordinary-matter density
$\rho_0$, in case $w_0 = 1$, and the ordinary mass $M_{(c)} = 4\pi\rho_0 R^3/(3c^2)$ 
a body with same $\rho_0$ and same central pressure $P_{(c)0}$ would have
in standard GR. 
Each curve represents a given value of $\sigma$ (indicated next to the corresponding curve).}
\label{fig:w01MMP0}
\end{figure}

The Eq.~(\ref{rhoncPcw0}) allows us to determine $\rho_\text{\it eff}$ and $P_\text{\it eff}$ explicitly in terms of 
$P_{(c)}$ [recall
Eqs.~(\ref{rhoeff}) and (\ref{Peff})], closing the system of Eqs.~(\ref{gr})-(\ref{meff}) for the functions
$g$, $\phi$, $P_{(c)}$, and $m_\text{\it eff}$
--- which should then be calculated numerically.
%, except for very special values 
%of $w_0$ and/or $\sigma$. 
One can easily verify that the special cases $w_0 = -1$ and $w_0 = 1/3$ turn out to be 
trivial, in the sense that, for these values,
$\rho_\text{\it eff} \equiv \rho_0$, which means that 
these models would be indistinguishable from standard GR. But this is certainly not the general 
behavior.
In Fig.~\ref{fig:w0m1by3diag}, we plot a mass-radius diagram for the case $w_0 = -1/3$, showing the possible
equilibrium configurations $(R, M_g)$ of objects with  radius $R$ and  
{\it effective} gravitational mass
 $M_g=m_\text{\it eff}(R)$, for different values of $\sigma$ and varying the central pressure
 $P_{(c)}(0) =: P_{(c)0}$.
 %from $0.05 \rho_0$ to $5\rho_0$. 
 The relevance of $M_g$ lies in the fact that
 this is the total mass anyone {\it assuming} GR
 would infer for the central object 
 through gravitational observations, even though its true total mass
 is given by $M= 4\pi\int_0^R dr\, r^2\rho$ and 
 the total mass in the form of ordinary matter is
 $M_{(c)} = 4\pi \rho_0 R^3/3$~\footnote{Obviously,  $M_{(c)}$ is 
 {\it not} 
 really
 the total mass $M_o$ in the form of ordinary matter;  the latter is obtained 
 by integrating $\rho_{(c)} = \rho_0$  using the proper-volume element:
 $M_o =4\pi \int_0^R dr\,r^2 \sqrt{g(r)}\,\rho_0$. 
 However, $M_{(c)}$ does give the 
 gravitational
 mass that would be inferred from standard GR for an object with proper density
 $\rho_0$ and radius $R$. Therefore, comparing $M_g$ with $M_{(c)}$
 is good enough for our purposes. }. 
 In Fig.~\ref{fig:w0m1by3MMR}, we plot
 the ratio 
 $ M_g/M_{(c)}$  as a function of the central pressure $P_{(c)0}$. We see
 that in this particular case $w_0 = -1/3$, an object will generally {\it appear} 
 to be more (resp., less)
 massive than it is in terms of ordinary (conservative) matter if $\sigma >0$
 (resp., $\sigma < 0$). In Fig.~\ref{fig:w0m1by3MMP0}, we  plot again the ratio
 $ M_g/M_{(c)}$  but now comparing $ M_g$ to
 the mass $M_{(c)}$ of an object with the {\it same central pressure} $P_{(c)0}$
in standard GR ($\sigma = 0$). Although central pressure is not a direct
observable, it is more directly related to the temperature of a star.
Therefore,
in spite of the fact that real stars are not really described by this highly
idealized constant-density model $\rho_{(c)} = \rho_0$,
it is interesting to note that the qualitative dependencies of 
$M_g/M_{(c)}$ with $\sigma$ 
in Figs.~\ref{fig:w0m1by3MMR} 
and \ref{fig:w0m1by3MMP0}
are inverted. Just for the sake of illustration, in
Figs.~\ref{fig:w01diag}-\ref{fig:w01MMP0} we plot the
same quantities as in Figs.~\ref{fig:w0m1by3diag}-\ref{fig:w0m1by3MMP0}, but
now for $w_0 = 1$. A peculiarity of the case $w_0 >0$ is that  
we cannot find equilibrium configurations for $P_{(c)0}$
arbitrarily large if $\sigma<0 $ (in case $0<w_0<1/3$) and
$\sigma>0 $ (in case $w_0>1/3)$.

\begin{figure}
\includegraphics[scale=0.55]{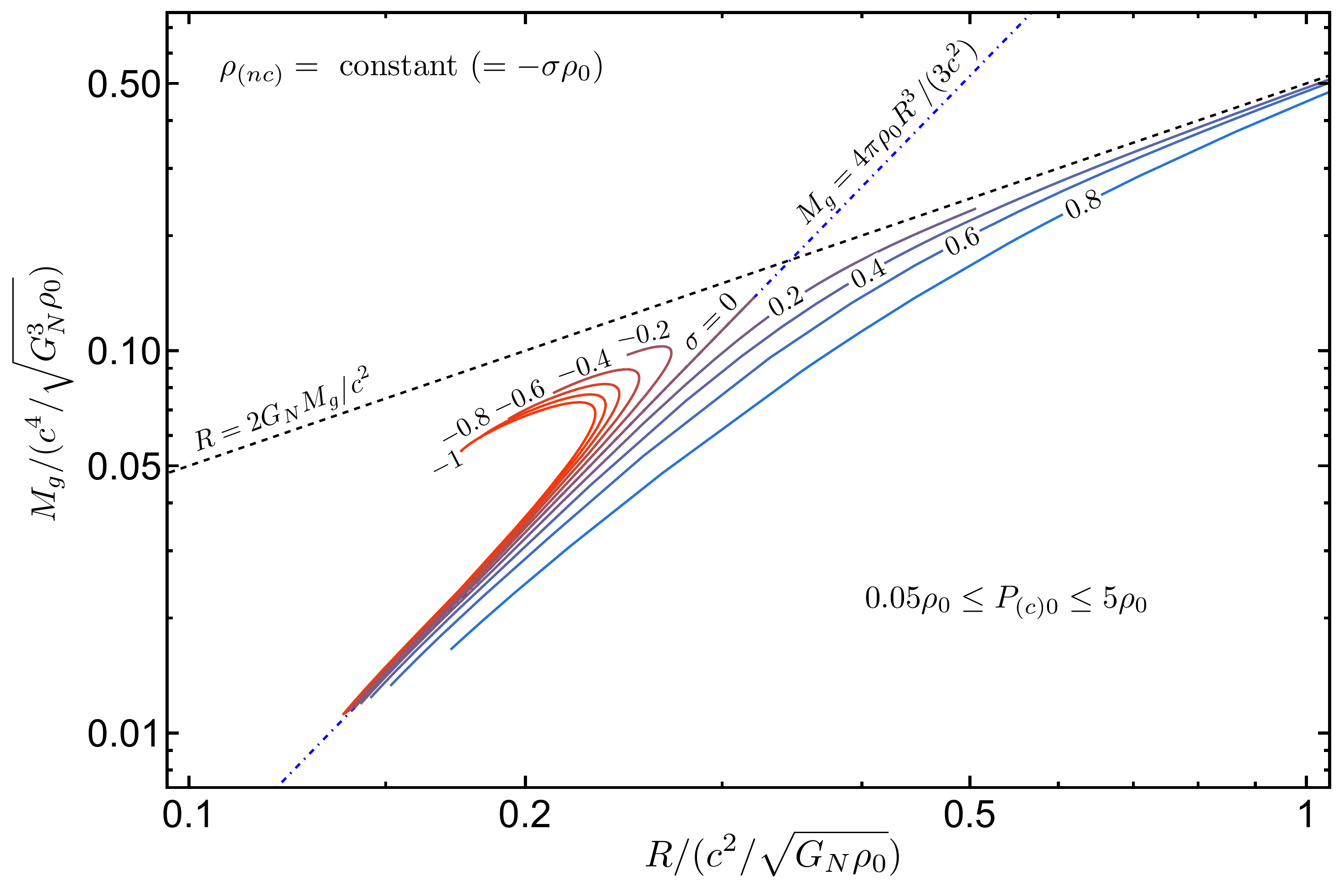}
\caption{Phase diagram showing the effective 
gravitational mass ($M_g$) and radius ($R$) of a spherical 
object with uniform energy densities $\rho_{(c)}=\rho_0$ and
$\rho_{(nc)} = -\sigma \rho_0$, 
in hydrostatic equilibrium. 
Each solid curve represents  equilibrium 
configurations
for values of 
ordinary-matter 
central pressure in the range $0.05 \rho_0< P_{(c)0}<5\rho_0$, 
for a given value
of $\sigma$ (indicated next to the corresponding curve). As expected, 
no equilibrium configuration is possible beyond the Schwarzschild limit
($R = 2G_N M_g/c^2$;  dashed line). The dot-dashed line gives the
ordinary mass contained in an object with radius $R$ and uniform energy 
density
$\rho_0$ --- 
which coincides with the gravitational mass an object with the same
density and radius would have (if possible) in standard GR.}
\label{fig:NR_diag}
\end{figure}

\begin{figure}
\includegraphics[scale=0.55]{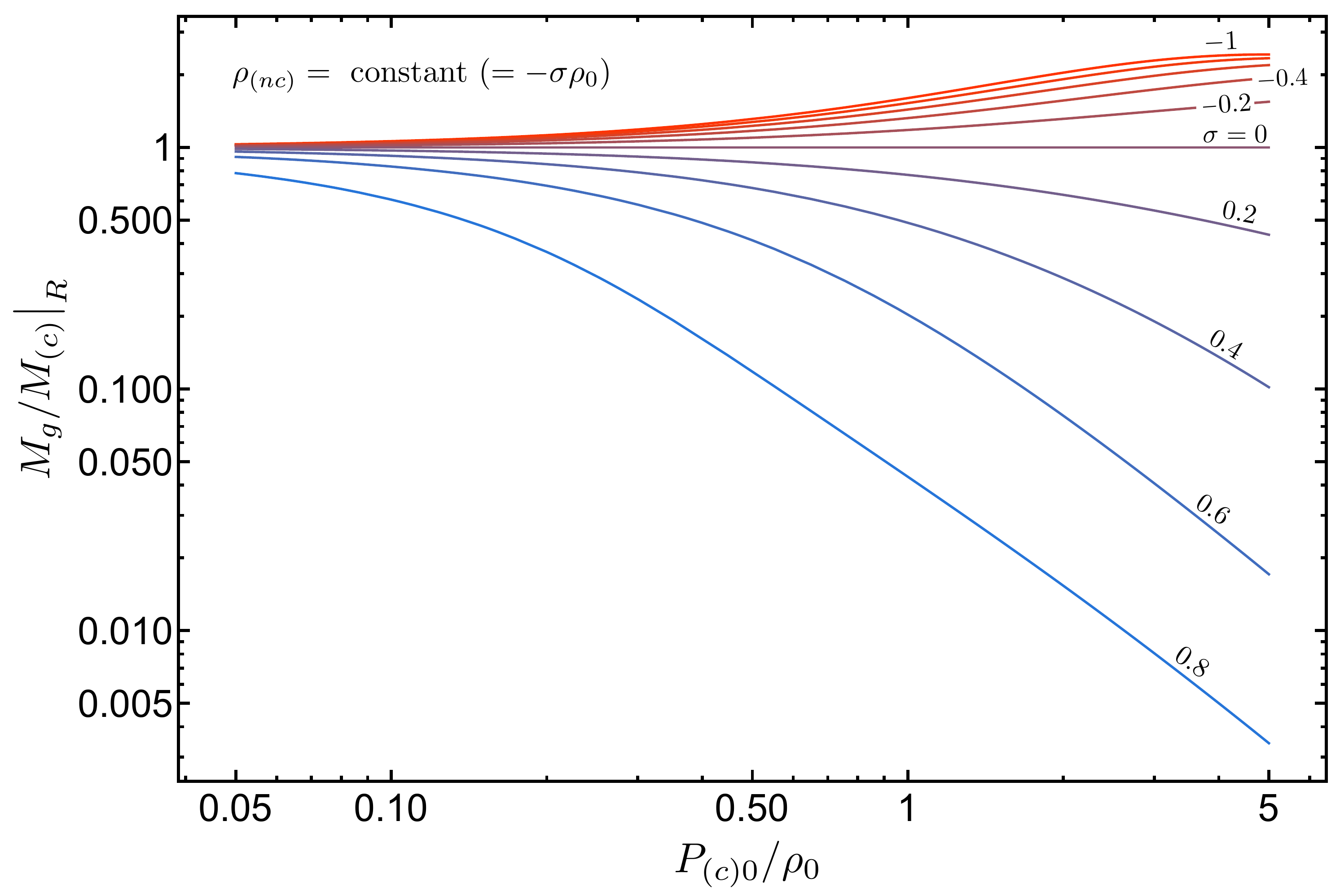}
\caption{Ratio between the effective gravitational mass $M_g$ of a central body with uniform 
energy densities
($\rho_{(c)} = \rho_0$, $\rho_{(nc)} = -\sigma \rho_0$)
and its ordinary mass $M_{(c)} = 4\pi\rho_0 R^3/(3c^2)$ --- the latter also being
the gravitational mass
a body with same $\rho_0$ and same radius $R$ would have (if possible)
in standard GR. Each curve represents a given value of $\sigma$ (indicated next to the corresponding curve, 
when possible; 
the curves in between the ones with $\sigma = -0.4$ and $\sigma = -1$
correspond to $\sigma =  -0.6$ and $-0.8$, in this order).}
\label{fig:NR_MMR}
\end{figure}

\begin{figure}
\includegraphics[scale=0.55]{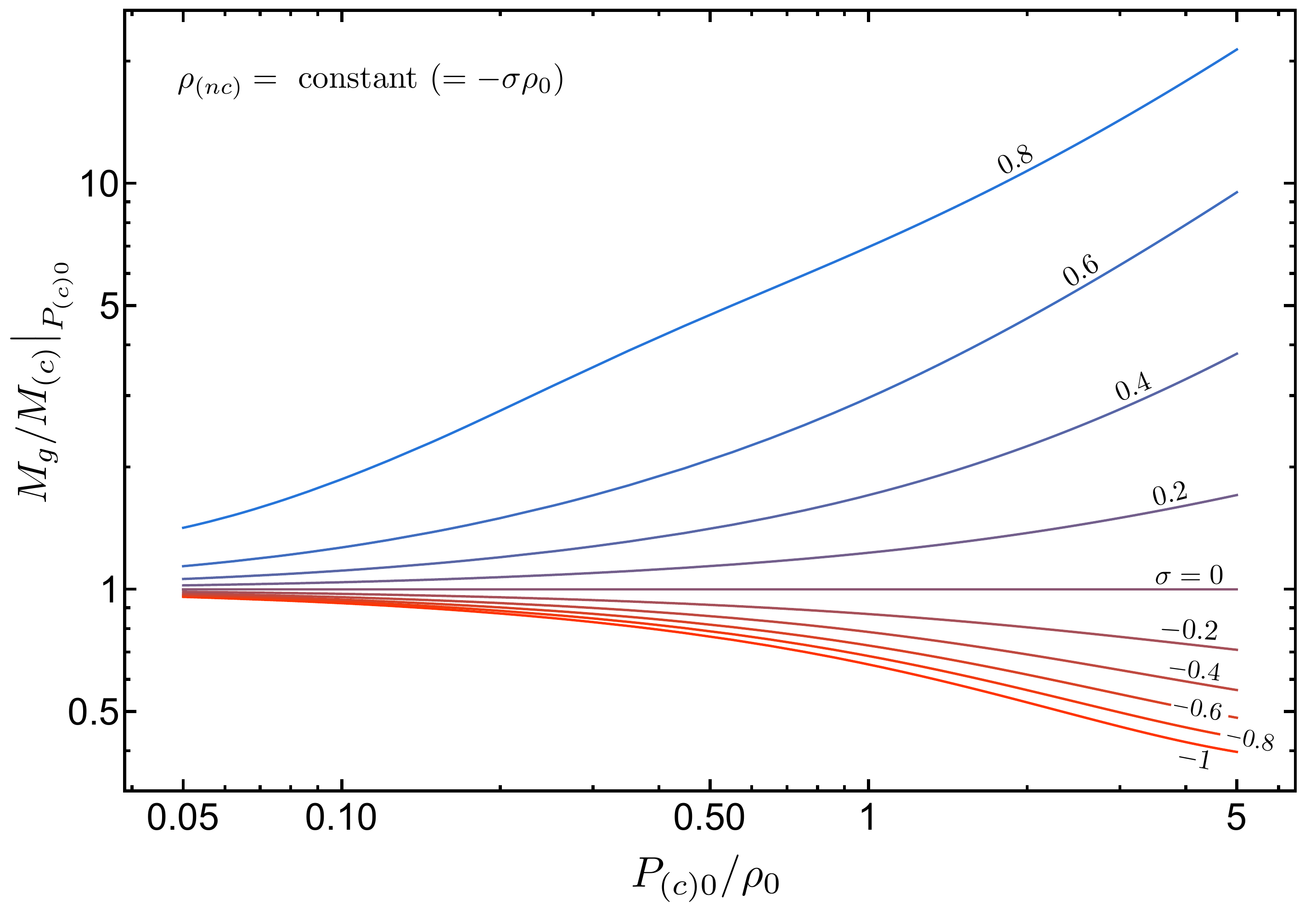}
\caption{Ratio between the effective gravitational mass $M_g$ of a central body with uniform 
energy densities
($\rho_{(c)}=\rho_0$, $\rho_{(nc)}=-\sigma \rho_0$) and the ordinary mass $M_{(c)} = 4\pi\rho_0 R^3/(3c^2)$ 
a body with same $\rho_0$ and same central pressure $P_{(c)0}$ would have
in standard GR. 
Each curve represents a given value of $\sigma$ (indicated next to the corresponding curve).}
\label{fig:NR_MMP0}
\end{figure}

\vskip 0.4cm
\noindent
%$\bullet$ 
{\bf (ii)}~{\bf{\underline{$\rho_{(nc)} = -\sigma \rho_{(c)} = -\sigma \rho_0$}:}}
\vskip 0.3cm
\noindent
Repeating the procedure described in the previous scenario, but now for 
$\rho_{(nc)} = -\sigma \rho_{(c)} = -\sigma \rho_0$
throughout the central object, we obtain that $P_{(nc)}$ satisfies
\begin{eqnarray}
(1-\sigma)\frac{dP_{(nc)}}{dr}
-(1-4\sigma)
\frac{P_{(nc)}}{(\rho_0+P_{(c)})} \frac{dP_{(c)}}{dr}
=
%\ln\left|1+\frac{P_{(c)}}{\rho_0}\right|
-\frac{[4\sigma(1-\sigma)\rho_0+3\sigma P_{(c)}]}{(\rho_0+P_{(c)})} 
\frac{dP_{(c)}}{dr},
\label{consNR}
\end{eqnarray}
whose solution subject to the boundary condition
$\left.P_{(nc)}\right|_R = -\left.\rho_{(nc)}\right|_R = \sigma \rho_0$
reads (for $\sigma \neq 1$~\footnote{For $\sigma = 1$ we get the 
uninteresting
solution $P_{(nc)} = -P_{(c)}$, 
which leads to $P_\text{\it eff} = \rho_\text{\it eff} = 0$.})
\begin{eqnarray}
P_{(nc)} = \sigma \rho_{0}-(\rho_0+P_{(c)})\left[1-
 \left(1+\frac{P_{(c)}}{\rho_0}\right)^{\!\!-\frac{3\sigma}{1-\sigma}}\right].
\label{PncNR}
\end{eqnarray}
Feeding this relation into Eqs.~(\ref{rhoeff}) and (\ref{Peff}) determines
$\rho_\text{\it eff}$ and $P_\text{\it eff}$ explicitly in terms of 
$P_{(c)}$, closing, again, the system of Eqs.~(\ref{gr})-(\ref{meff}) for
$g$, $\phi$, $P_{(c)}$, and $m_\text{\it eff}$.
In Figs.~\ref{fig:NR_diag}-\ref{fig:NR_MMP0} we plot the same quantities as
for the previous scenario (except for the omission of the uninteresting case
$\sigma = 1$). Notice that the effect of different values of $\sigma$ in 
the present scenario exhibits some similarities with the case $w_0>1/3$ of the
previous scenario.

\section{Modified cosmology}
\label{sec:modcosm}

Now, we proceed to investigate possible consequences of Eq.~(\ref{modEqG}) 
to spatially homogeneous and isotropic cosmology. Considering the 
FLRW line element
$$
ds^2 = -c^2dt^2+a(t)^2d\Sigma_{(k)}^2,
$$
with $d\Sigma_{(k)}^2$ being the Riemannian line element of a unit
3-sphere ($k = 1$), an Euclidean 3-space ($k = 0$) or a unit hyperbolic 3-space ($k = -1$) and
$a(t)$ being the scale factor, Eq.~(\ref{modEqG}) implies the usual Friedmann equations for the 
effective energy density and pressure [see Eqs.~(\ref{rhoeff}) and (\ref{Peff})]:
\begin{eqnarray}
\left(\frac{\dot{a}}{a}\right)^{\!\!2}\!\! &=& \frac{8\pi G_N }{3c^2}\rho_\text{\it eff} -\frac{kc^2}{a^2},
\label{FriedEqH}
\\
\frac{\ddot{a}}{a} &=& -\frac{4\pi G_N}{3c^2}(\rho_\text{\it eff}-3P_\text{\it eff}),
\label{FriedEqq}
\end{eqnarray}
where each dot represents derivative w.r.t.~$t$. Moreover, Eq.~(\ref{model1j}) applied to each constituent 
${\cal U}_j$ (with no nongravitational interactions)
reads
\begin{eqnarray}
\frac{d\rho_j}{dt} + 3\frac{\dot{a}}{a}(\rho_j+P_j) = -\sigma_j\frac{d}{dt}(\rho_\text{\it eff}-3P_\text{\it eff}),
\label{FLRWconsj}
\end{eqnarray}
which summed over $j$ leads to
\begin{eqnarray}
\frac{d\rho_\text{\it eff}}{dt} + 3\frac{\dot{a}}{a}(\rho_\text{\it eff}+P_\text{\it eff}) = 0
\label{FLRWconseff}
\end{eqnarray}
--- the latter also being a direct consequence of Eqs.~(\ref{FriedEqH}) and (\ref{FriedEqq}). In terms of
${\cal U}_{(c)}$ and ${\cal U}_{(nc)}$, these equations imply
\begin{eqnarray}
\frac{d}{dt}\left[\rho_{(nc)}-3\bar{\sigma}(\rho_{(nc)}+P_{(nc)})\right]+3(1-4\bar{\sigma})
\frac{\dot{a}}{a}(\rho_{(nc)}+P_{(nc)})
=
-\bar{\sigma}\frac{d}{dt}(\rho_{(c)}-3P_{(c)}).
\label{consNC_C}
\end{eqnarray}
It is interesting to note that out of the usual ingredients considered in cosmological scenarios
--- namely, radiation, whose energy density and pressure are, respectively,
$\rho_r =\rho_{r0}(a_0/a)^4 $ and $P_r = \rho_r/3$, and
nonrelativistic matter (``dust''), whose energy density and pressure are $\rho_m =\rho_{m0}(a_0/a)^3 $ and
 $P_m = 0$ (subscript ``0'' indicates present value of the corresponding quantity) ---,
only nonrelativistic matter  contributes to the r.h.s.\ of Eq.~(\ref{consNC_C}).

Assuming, for simplicity, that 
${\cal U}_{(nc)}$ is characterized by a constant equation of state $w_{(nc)} 
:= P_{(nc)}/\rho_{(nc)}$
and using $\rho_{(c)} = \rho_{m} +\rho_{r}$ and $P_{(c)} = P_{r}+P_{m}$, 
Eq.~(\ref{consNC_C}) can be integrated to give
\begin{eqnarray}
\rho_{(nc)} 
= 
\frac{\bar{\sigma} \rho_{m}}{[w_{(nc)}-\bar{\sigma}(1+w_{(nc)}) ]} +C a^{-\frac{3(1-4\bar{\sigma})(1+w_{(nc)})}{[1-3\bar{\sigma}(1+w_{(nc)})] }} ,
\label{FLRWrhonc}
\end{eqnarray}
where $C$ is a mere integration constant. 
%which we shall discuss later.
Substituting this result into the expression for $\rho_\text{\it eff}$
--- which is the relevant quantity to determine the expansion rate of the 
universe [see Eq.~(\ref{FriedEqH})] --- and $P_\text{\it eff}$ --- which influences the
acceleration of the expansion [see Eq.~(\ref{FriedEqq})] ---, we have
\begin{eqnarray}
\rho_\text{\it eff} 
&=& 
\frac{w_{(nc)} \rho_{m}}{\alpha[w_{(nc)}-\bar{\sigma}(1+w_{(nc)}) ]}+ \frac{\rho_r}{\alpha}+\frac{[1-3\bar{\sigma}(1+w_{(nc)})]}{\alpha (1-4\bar{\sigma})}C a^{-\frac{3(1-4\bar{\sigma})(1+w_{(nc)})}
{[1-3\bar{\sigma}(1+w_{(nc)})] }} ,
\label{FLRWrhoeff}
\\
P_\text{\it eff} 
&=& 
 \frac{\rho_r}{3\alpha}+\frac{[w_{(nc)}-\bar{\sigma}(1+w_{(nc)})]}{\alpha (1-4\bar{\sigma})}C a^{-\frac{3(1-4\bar{\sigma})(1+w_{(nc)})}
{[1-3\bar{\sigma}(1+w_{(nc)})] }} .
\label{FLRWPeff}
\end{eqnarray}

Now, we must proceed with some caution. The reader might be ready to 
identify $w_{(nc)}\equiv w_0$ and use Eq.~(\ref{alpha}), which would then 
lead to
\begin{eqnarray}
\rho_\text{\it eff} 
&=& 
\rho_{m}+ \frac{\rho_r}{1+\sigma(1+w_0^{-1})}+
\frac{[1+\sigma (1+w_0)(w_0^{-1}-3)]}{[1+\sigma(1+w_0^{-1})][ 1+\sigma(w_0^{-1}-3)]}C a^{-\frac{3(1+w_0)[ 1+\sigma(w_0^{-1}-3)]}
{[1+\sigma (1+w_0)(w_0^{-1}-3)] }} ,
\label{FLRWrhoeffwncw0}
\\
P_\text{\it eff} 
&=& 
\frac{\rho_r}{3[1+\sigma(1+w_0^{-1})]}+\frac{w_0}{[1+\sigma(1+w_0^{-1})][ 1+\sigma(w_0^{-1}-3)]}C a^{-\frac{3(1+w_0)[ 1+\sigma(w_0^{-1}-3)]}
{[1+\sigma (1+w_0)(w_0^{-1}-3)] }}.
\label{FLRWPeffwncw0}
\end{eqnarray}
And this indeed makes sense in case ${\cal U}_{(nc)}$ has a single component
with a fixed equation of state
--- as considered in scenario (i) of the previous section. However, in case
 $\rho_{(nc)}$ and $P_{(nc)}$ can be sourced independently --- as in scenario
 (ii) of the previous section ---, $w_{(nc)}$ and $w_0$ do not have to match.
 In fact, notice that $w_0$, referring to the static, spatially 
 inhomogeneous energy distribution in the Newtonian regime, cannot vanish,
 while
 in the spatially homogeneous scenario nothing forbids $P_{(nc)}$ 
 (and, therefore, $w_{(nc)}$) to be null. In particular, a simple cosmological
 analogous of scenario (ii) of the previous section
 %adapted to the cosmological
 %context 
 would be one with $w_0 = -1$ (see Sec.~\ref{sec:NRw}) and
 $w_{(nc)} = 0$ (a dust-like homogeneous dark constituent), leading to
 \begin{eqnarray}
\rho_\text{\it eff} 
= 
\rho_r+
\frac{(1-3\sigma)}{(1-4\sigma)}C a^{-\frac{3(1-4\sigma)}{(1-3\sigma)}} ,
\label{FLRWrhoeffwNR}
\\
P_\text{\it eff} 
= 
\frac{\rho_r}{3}-
\frac{\sigma}{(1-4\sigma)}C a^{-\frac{3(1-4\sigma)}{(1-3\sigma)}} .
\label{FLRWPeffwNR}
\end{eqnarray}

The most interesting aspect of the results presented in Eqs.~(\ref{FLRWrhoeff})-(\ref{FLRWPeffwNR}) is that 
a nonconservative constituent ${\cal U}_{(nc)}$ 
with given equation of state $w_{(nc)}$ can  {\it mimic} a conservative 
ingredient with an effective equation of state 
\begin{eqnarray}
w_\text{\it eff} = \frac{w_{(nc)}-\bar{\sigma} (1+w_{(nc)})}
{1-3\bar{\sigma} (1+w_{(nc)})}
\label{FLRWweffsbarwnc}
\end{eqnarray}
when the data are interpreted through the lens of 
standard GR. In particular, if we consider $w_{(nc)} = w_0$ 
[as in Eqs.~(\ref{FLRWrhoeffwncw0}) and (\ref{FLRWPeffwncw0})],
we have
\begin{eqnarray}
w_\text{\it eff} = \frac{w_0}
{1+{\sigma} (1+w_{0})(w_0^{-1}-3)}.
\label{FLRWweffsw0}
\end{eqnarray}
Moreover, in the scenario where $w_{(nc)}$ and $w_0$ are independent, 
there is also a residual effect of the nonconservative
equations on the conservative constituents ${\cal U}_{(c)}$: interpreted
through the lens of standard GR, the effective 
amount of nonrelativistic matter and radiation relate to the actual values
by the multiplicative factors
$\alpha^{-1}w_{(nc)}/[w_{(nc)}-\bar{\sigma}(1+w_{(nc)})]$ and 
$\alpha^{-1}$, respectively [see Eq.~(\ref{FLRWrhoeff})]. All these results combined open up the possibility of a
``dark'' ingredient ${\cal U}_{(nc)}$ being responsible, at once, for driving 
the cosmic expansion through an accelerating phase (if $w_\text{\it eff}<-1/3$, $\rho_\text{\it eff}>0$) and 
for an excess in the (gravitationally) observed
amount of nonrelativistic matter 
(if $\alpha^{-1}w_{(nc)}/[w_{(nc)}-\bar{\sigma}(1+w_{(nc)})]>1$).

\begin{figure}
\includegraphics[scale=0.55]{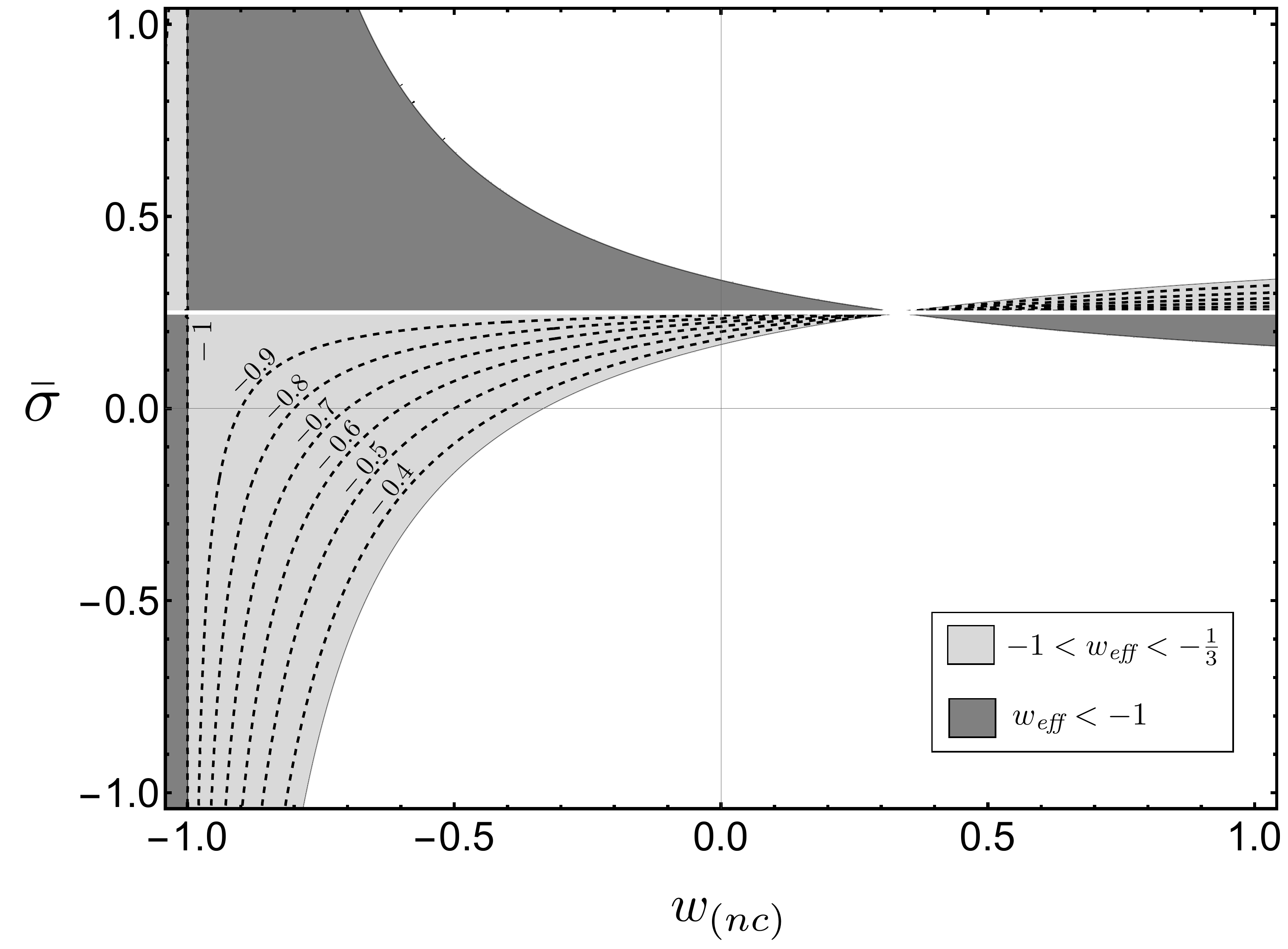}
\caption{Diagram showing the values $(w_{(nc)},\bar{\sigma})$ for which 
$-1<w_\text{\it eff}<-1/3$ (light-gray region) and $w_\text{\it eff}<-1$ (dark-gray region) --- 
according to Eq.~(\ref{FLRWweffsbarwnc}). The dashed curves represent constant values of $w_\text{\it eff}$
(given  next to the corresponding curve).}
\label{fig:diagsbarwnc}
\end{figure}

\begin{figure}
\includegraphics[scale=0.55]{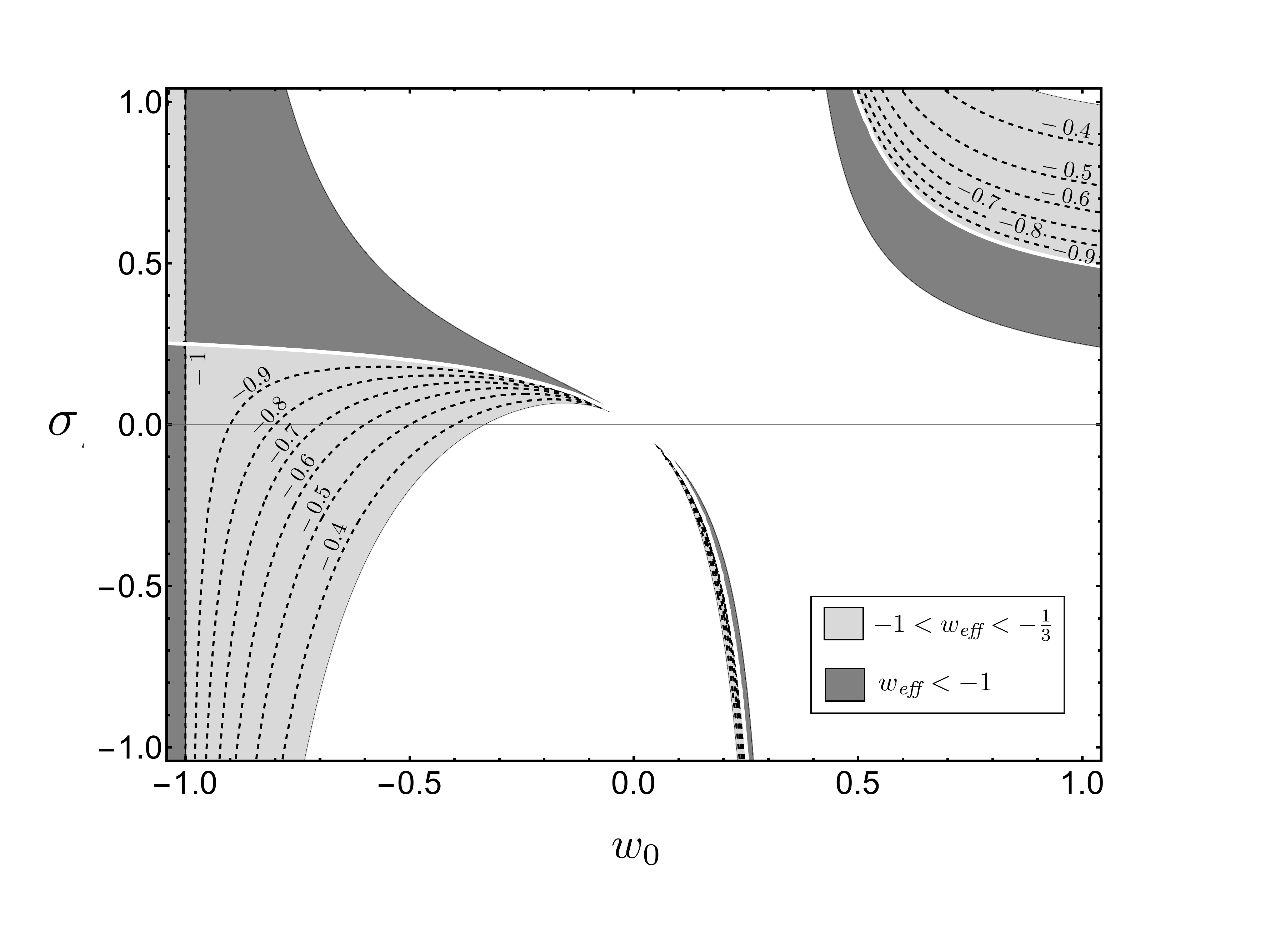}
\caption{Diagram showing the values $(w_0,{\sigma})$ for which 
$-1<w_\text{\it eff}<-1/3$ (light-gray region) and $w_\text{\it eff}<-1$ (dark-gray region) --- 
according to Eq.~(\ref{FLRWweffsw0}). The dashed curves represent constant values of $w_\text{\it eff}$
(given  next to the corresponding curve). }
\label{fig:diagsw0}
\end{figure}

Using Eqs.~(\ref{FLRWweffsbarwnc}) 
and (\ref{FLRWweffsw0}), in Figs.~\ref{fig:diagsbarwnc} and \ref{fig:diagsw0} we highlight the regions in the 
parameter spaces $\{(w_{(nc)},\bar{\sigma})\}$ and
$\{(w_{0},{\sigma})\}$, respectively, where $w_\text{\it eff}<-1/3$ --- which represent models with late
accelerating cosmic expansion (for $\rho_\text{\it eff}>0$). We also
represent  curves where $w_\text{\it eff}$ assumes some constant values 
(given next to the corresponding curve). This is only to make it clearer 
that in the scenario where
$w_{(nc)}$ and $w_0$ are independent (Fig.~\ref{fig:diagsbarwnc}), 
even a dust-like nonconservative constituent ($w_{(nc)} = 0$) could,
in principle, mimic a dark-energy-like ingredient in standard GR.

%Some remarks must be made regarding the integration constant $C$ appearing in the 
%equations above.
%First, note that $C$ is really constant only as long as there is no energy exchange
%among the constituents of ${\cal U}_{(c)}$ --- in this case, ordinary matter and 
%radiation.

\section{Conclusion}
\label{sec:final}

In the present work, we have investigated to which extent the
canon of local energy-momentum conservation applied 
%separately 
to matter/interaction fields (other than gravity) 
--- as enforced by standard GR and other 
metric theories of gravity through Eq.~(\ref{eq:metrictheories}) with 
$\nabla_a{\cal G}^{ab}\equiv 0$ --- can be challenged to allow local energy-momentum
exchange between the nongravitational system and the spacetime. There are several
ways in which such an exchange of energy-momentum can take place without violating 
the
other (seemingly more natural) canon of general covariance. As a proof of concept and in the
absence of a fundamental description of how such an exchange could occur, here we adopted 
a phenomenological approach and
explored the simplest model one could devise, given by 
Eqs.~(\ref{model1}) and (\ref{model1j}) --- 
which generalizes the model proposed by Rastall fifty years ago~\cite{Rastall}, 
explored in subsequent
works~\cite{LH,BDFPR,OVFC,OVF,Visser}. 
Although the distinction between Rastall's theory --- Eq.~(\ref{model1}) with an ingredient-independent
parameter $\sigma$ ---
and our Rastall-like model --- Eq.~(\ref{model1j}) with ingredient-dependent
parameters $\sigma_j$ --- might seem unimportant at first, 
we have explicitly shown that it is enough, in a universe containing ``dark'' ingredients
(such as ours), to render the latter immune to the criticisms
which are used to discard the former as a viable model.
%~\cite{LH}.
%Hence, even though
%We have then shown 
%how the Newtonian regime can be recovered  in this context (Sec.~\ref{sec:modeq})
%--- which seems to have been overlooked in the literature ---
%and also how
%the Schwarzschild and the FLRW solutions would be affected in this particular model
%(Secs.~\ref{sec:modSch} and \ref{sec:modcosm}, respectively). 
Interestingly enough, in spite of
%although 
our primary motivation
%for considering local energy-momentum exchange between the spacetime and
%matter/interaction fields 
relying purely on ``naturality'' arguments, we showed that this issue of ``nonconservative gravity'' 
gets
 tied to the existence of ``dark'' ingredients 
 in our Universe --- a connection which seems to 
have
been overlooked in the literature.
%--- i.e., those we only 
%perceive through their gravity effects. 
This fact is essential when confronting the consequences of the model with observations and it
%seems to have been missed in the literature  and it 
bears effects on many of the results
presented in this manuscript (summarized in Figs.~\ref{fig:w0m1by3diag}-\ref{fig:diagsw0}).
In fact, were we living in a Universe with no room for 
``dark'' ingredients and the Rastall-like model
given by Eq.~(\ref{model1j}) (with $T^{ab}_{(j)}$ 
being the physical stress-energy-momentum tensor)
would have been discarded (or at least have the $\sigma_j$ parameters severely constrained) by simple
observations of lower bounds on pressure-density ratios in the Newtonian regime (Sec.~\ref{sec:modeq}). 
On the other hand,  as we have discussed in Sec.~\ref{sec:modcosm}, 
 a ``dark'' nonconservative ingredient
satisfying Eq.~(\ref{model1j}) could, in principle, mimic dark-energy and dark-matter effects, at once, when
observations are interpreted 
through the lens of standard GR. Hence, although a violation
of $\nabla_a T^{ab} = 0$ may seem ``radical'' at first, analyses like the one
presented here suggest otherwise: not only it is natural (from a logical perspective), but it also
finds  room to be true in our observed Universe.

%\acknowledgments

\end{document}